\documentclass[5p,authoryear]{elsarticle}

\usepackage[utf8]{inputenc} % allow utf-8 input
\usepackage[T1]{fontenc}    % use 8-bit T1 fonts
\usepackage{lmodern}
\usepackage{hyperref}       % hyperlinks
\usepackage{url}            % simple URL typesetting
\usepackage{booktabs}       % professional-quality tables
\usepackage{amsfonts}       % blackboard math symbols
\usepackage{nicefrac}       % compact symbols for 1/2, etc.
\usepackage{microtype}      % microtypography
\usepackage{amsmath}
\usepackage{bm}
\usepackage{multirow}
\usepackage{color}
\usepackage{flushend}
\usepackage{graphicx}

\usepackage{caption}
\def\tx#1{{\tt\small{#1}}}

\newcommand\numberthis{\addtocounter{equation}{1}\tag{\theequation}}

\graphicspath{{./figures/}}
\DeclareGraphicsExtensions{.pdf,.jpg,.png}

\usepackage{xcolor}
\usepackage{soul}
\usepackage[normalem]{ulem}

\definecolor{rosepale}{rgb}{1.0, 0.7, 1.0}
\definecolor{rougepale}{rgb}{1.0, 0.4, 0.4}
\definecolor{violetpale}{rgb}{0.8, 0.4, 1.0}
\definecolor{jaunepale}{rgb}{0.8, 0.8, .2}
\definecolor{cyanpale}{rgb}{0.0, 0.8, .8}

\title{A Variational Prosody Model for Mapping the Context-Sensitive Variation of Functional Prosodic Prototypes}

\author[feeit,gipsa]{Branislav~Gerazov}
\ead{gerazov@feit.ukim.edu.mk}

\author[gipsa]{G\'erard~Bailly}
\ead{gerard.bailly@gipsa-lab.fr}

\author[gipsa]{Omar~Mohammed}
\ead{omar-samir.mohammed@gipsa-lab.fr}

\author[ucl]{Yi~Xu}
\ead{yi.xu@ucl.ac.uk}

\author[idiap]{and Philip~N.~Garner}
\ead{Phil.Garner@idiap.ch}

\address[feeit]{%
  Faculty of Electrical Engineering and Information Technologies,
  Ss. Cyril and Methodius University in Skopje,
  Skopje, Macedonia
}
\address[gipsa]{%
  Univ. Grenoble-Alpes, CNRS, Grenoble-INP,
  GIPSA-lab,
  Grenoble, France
}
\address[ucl]{%
  Department of Speech, Hearing and Phonetic Sciences,
  University College London,
  London, UK
}
\address[idiap]{%
  Idiap Research Institute,
  Martigny, Switzerland
}

\begin{document}

\begin{abstract}
  The quest for comprehensive generative models of intonation that link linguistic and paralinguistic functions to prosodic forms has been a longstanding challenge of speech communication research. Traditional intonation models have given way to the overwhelming performance of deep learning (DL) techniques for training general purpose end-to-end mappings using millions of tunable parameters. The shift towards black box machine learning models has nonetheless posed the reverse problem -- a compelling need to discover knowledge, to explain, visualise and interpret. Our work bridges between a comprehensive generative model of intonation and state-of-the-art DL techniques. We build upon the modelling paradigm of the Superposition of Functional Contours (SFC) model and propose a Variational Prosody Model (VPM) that uses a network of variational contour generators to capture the context-sensitive variation of the constituent elementary prosodic contours. We show that the VPM can give insight into the intrinsic variability of these prosodic prototypes through learning a meaningful prosodic latent space representation structure. We also show that the VPM is able to capture prosodic phenomena that have multiple dimensions of context based variability. Since it is based on the principle of superposition, the VPM does not necessitate the use of specially crafted corpora for the analysis, opening up the possibilities of using big data for prosody analysis.
  In a speech synthesis scenario, the model can be used to generate a dynamic and natural prosody contour that is devoid of averaging effects.
\end{abstract}

\maketitle
\section{Introduction}

The quest for comprehensive generative models of intonation that link linguistic and paralinguistic functions to prosodic forms has been a longstanding challenge of speech communication research. The first quantitative models proposed in the early 70s by \cite{fujisaki1971generative},
% G{\aa}rding
\cite{gaarding1983generative} and consorts, the transcription frameworks for intonation and rhythm, e.g. TOBI~\citep{silverman1992tobi}, INTSINT~\citep{hirst2005form}, etc., and the more complex generative models of intonation originally proposed at the turn of the century by \cite{bailly2005sfc}
and \cite{xu2005speech}, were focusing on extracting elementary prosodic atoms or \emph{geons} \citep{biederman1987recognition} that syntactically combine to form the observed prosodic contours. The quest for building blocks and organisation principles of intonation systems is fundamental for understanding how oral languages are built, learnt and evolve. Modelling constraints also spare data requirements -- we certainly experience few samples of the extraordinary varieties of expressive emotions, e.g. the \cite{baron2004mind} taxonomy comprises 412 auditory-visual patterns, but we master most of them on a large arsenal of textual supports.

Starting  at the turn of the century, within the paradigm of statistical parametric text-to-speech (TTS) synthesis, explicit prosody models gave way to machine learning techniques based on Hidden Markov Models \citep{yoshimura1999simultaneous, obin2011stylization}
and later Deep Learning (DL) \citep{zen2013statistical, zen2015acoustic, yin2016modeling, wang2017rnn}.
Today, state-of-the-art TTS systems are built on top of the overwhelming performance of DL techniques for training general purpose end-to-end mappings using millions of tunable parameters \citep{wang2017tacotron, arik2017deep, sotelo2017char2wav, taigman2018voiceloop, shen2018natural}.
The success of these black box machine learning models has nonetheless now posed the reverse problem -- a compelling need to discover knowledge, to explain, visualise, and interpret.
For example, one of the key challenges is the disentanglement of style from textual content and its control \citep{wang2018style, skerry-ryan2018towards, hsu2018hierarchical, ma2018generative}.
% , robinson2019sequence}.

On the other side, there is still a need to identify, analyse and catalogue the immense variability of prosody in language, even in well studied languages like English \citep{goodhue2016toward}.
This necessitates the creation of specially tailored corpora that can be tedious to design and create, especially if the prosodic variability depends on the interaction between multiple linguistic and paralinguistic dimensions \citep{liu2005parallel}.
%\citep{wagner2018effect}.

Our work bridges between a comprehensive generative model of intonation and state-of-the-art DL techniques. We build upon the modelling paradigm of the Superposition of Functional Contours (SFC) model \citep{bailly2005sfc, bailly2002learning, morlec2001generating}, by designing a deep neural network architecture \citep{goodfellow2016deep} that incorporates variational encoding \citep{kingma2013auto, zhao2017infovae} and is jointly trained using backpropagation \citep{kingma2014adam}.
Our proposed Variational Prosody Model (VPM) has the capability of capturing context-sensitive variations of the constituent elementary prosodic atoms, i.e. multiparametric prosodic prototypes, or \emph{clich{\'e}s} \citep{fonagy1983cliches} that encode given linguistic and paralinguistic functions onto units of variable size, or scope.
These prototypes are multiparametric because they feed a multiparametric prosodic score that includes intonation, rhythm, but can also include function-specific eye, head and body movements, etc.
Moreover, the VPM is able to learn a meaningfully structured prosodic latent space representation of this variability.

We compare here the modelling power of the VPM to the original SFC and the weighted SFC (WSFC)~\citep{gerazov2018wsfc}, which adjoins a weighting module to each contour generator for scaling its contribution in the decomposition, as well as a baseline deep model based on the Merlin speech synthesiser prosody module \cite{wu2016merlin}.
We demonstrate that the proposed deep architecture of the VPM while bringing added value, gives comparable if not better modelling performance than our previous models.
Finally, we show that within the function-specific prosodic latent spaces, the VPM is able to capture spatiotemporal variations of the \emph{clich{\'e}s} that go beyond modelling their mean (SFC) or simple amplitude scaling (WSFC).
This can be used in the exploration of context-specific prosodic phenomena. Moreover, the fact that it is based on decomposing prosody into its constituents, allows the use of the VPM on large unconstrained natural speech data, eliminating the need of specially designed corpora.
We also believe this variational modelling power of the VPM can readily be used in a speech synthesis scenario to generate rich and natural sounding prosodic contours.

\section{The SFC and WSFC}

\begin{figure}[t]
  \setlength\belowcaptionskip{-10pt}
    \centering
	\includegraphics[width=.9\columnwidth]{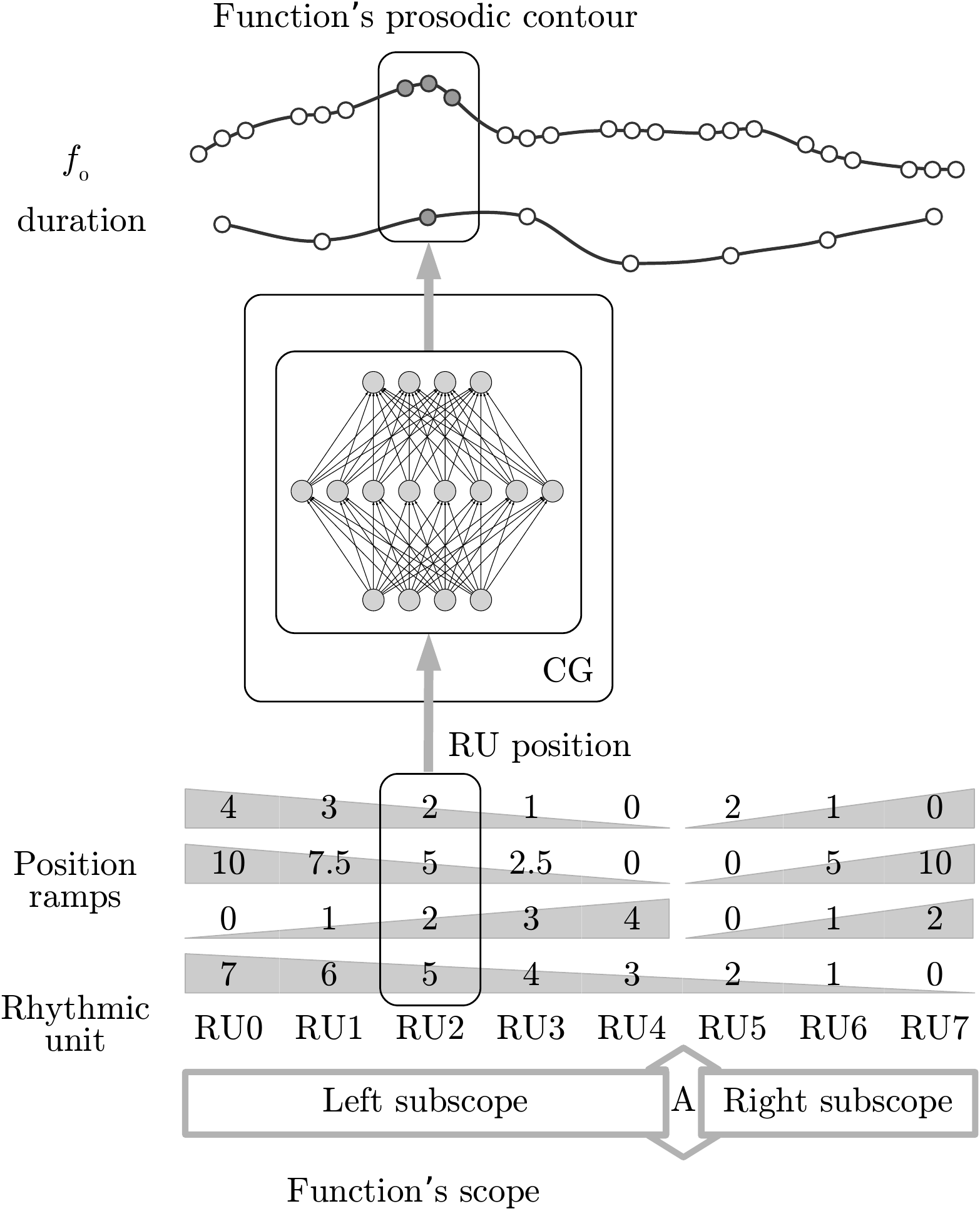}
      \caption{The SFC contour generator (CG) generating a function's prosody contour for a given scope one rhythmic unit at a time. The shown function has two subscopes divided by an anchor point (A).}
	\label{fig:sfc}
\end{figure}

The SFC modelling paradigm supposes that multiple functions acting on multiple units are transmitted via prosody using a simple channel sharing procedure. Thus, the prosody of the utterance is simply performed by overlapping-and-adding all contributing multiparametric prototypes.
The problem of decomposing prosody into these elementary patterns is ill-posed because the SFC does not impose any a priori constraints on the spatiotemporal patterns such as bandwidth or shape. In the SFC, these \emph{clich{\'e}s}, in fact, emerge from statistical modelling -- an iterative analysis-by-synthesis training process is used to train function-specific contour generators (CGs)~\citep{morlec1997phd}, shown in Fig.~\ref{fig:sfc}.
The SFC has been successfully used to model different functions acting at various linguistic levels, including: attitudes~\citep{morlec2001generating}, grammatical dependencies~\citep{morlec1998evaluating}, cliticisation~\citep{bailly2002learning}, focus~\citep{brichet2004prosodie}, as well as tones in Mandarin~\citep{chen2004superposed}.

Each of the contour generators is trained to encode a specific functional contour across different scopes. The scope designates the number of rhythmic units that the contour spans, e.g. syllables or inter perceptual centre groups~\citep{campbell1992syllable}.
In the SFC, the CGs are implemented as a single shallow neural network that takes as input the absolute and relative position of the current RU in the linguistic function's scope w.r.t. its boundaries and anchor point, encoded through input ramps shown in Fig.~\ref{fig:sfc}. Based on this input, the CG then outputs the prosodic contour for the function, one RU at a time.
The prosodic contour comprises pitch targets for each RU's vowel nucleus, and a duration modification coefficient based on the average RU duration that solely depends on its phonetic constituents \citep{bailly2005sfc}.

\begin{figure*}[th]
  \setlength\belowcaptionskip{-10pt}
 \centering
%     \begin{minipage}[t]{.48\linewidth}
\centering
%  \hfill
%  \includegraphics[height=5.15cm]{wsfc_cg_v2.pdf}
 \includegraphics[width=.22\linewidth]{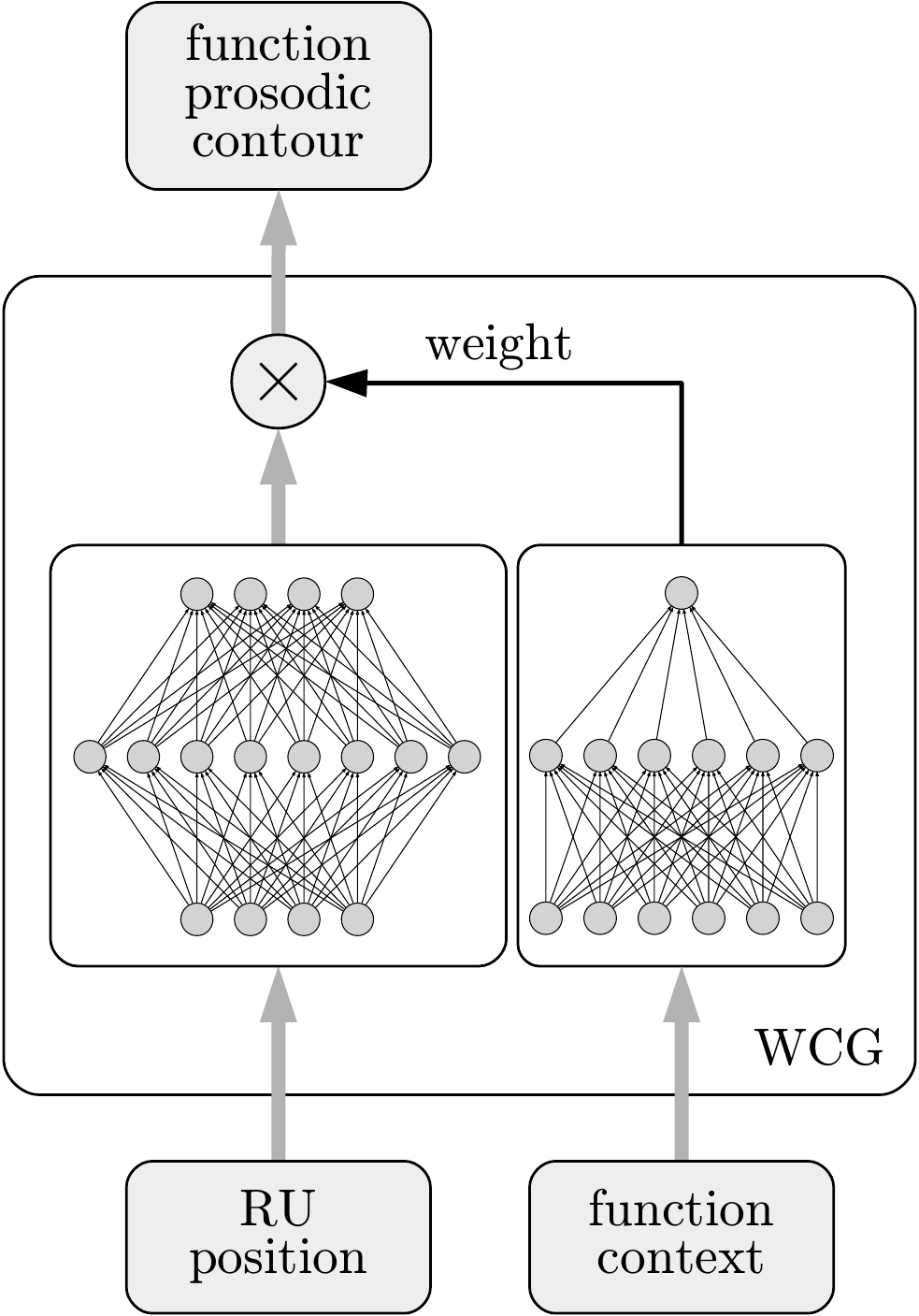}
 \hfill
 \includegraphics[width=.21\linewidth]{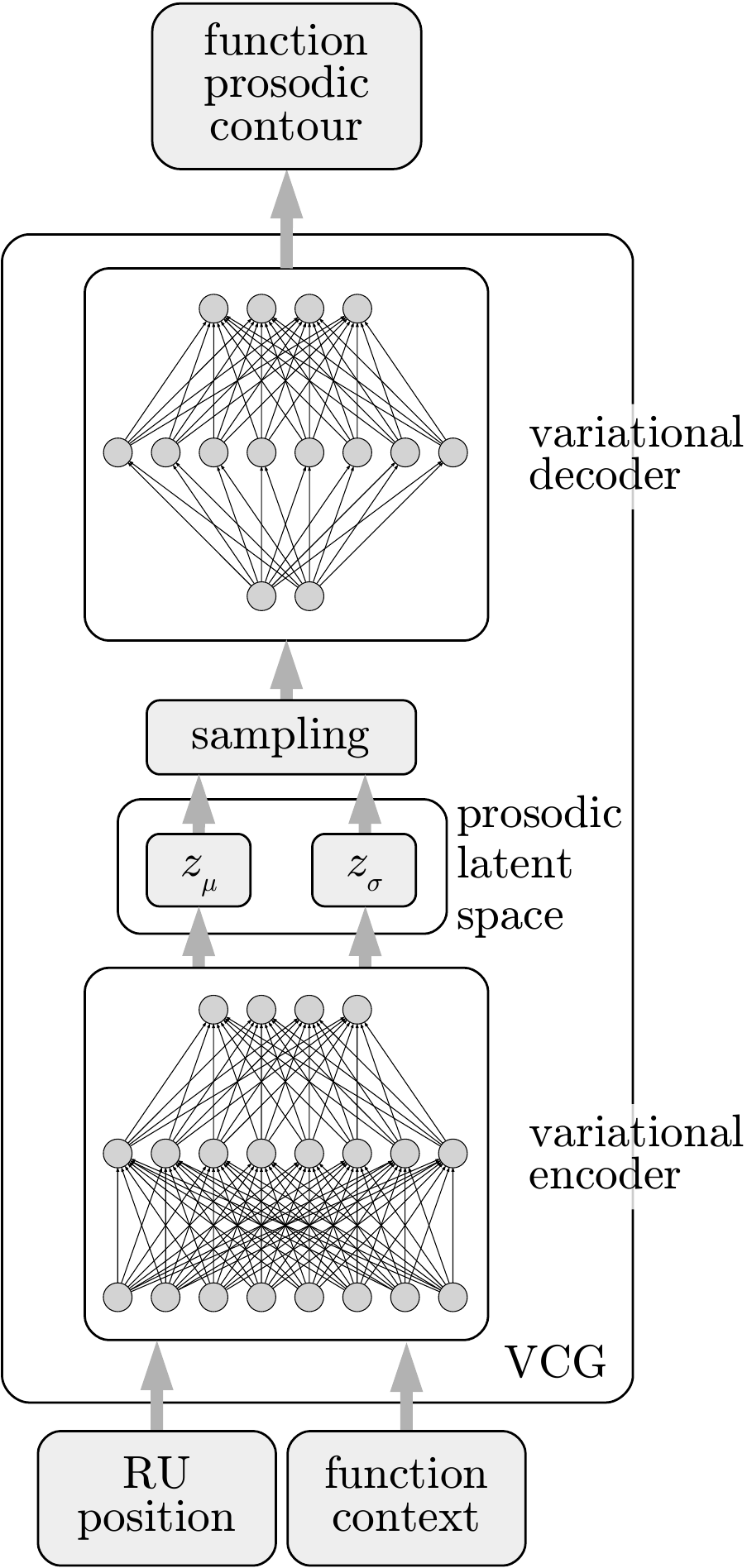}
 \hfill
 \includegraphics[width=.48\linewidth]{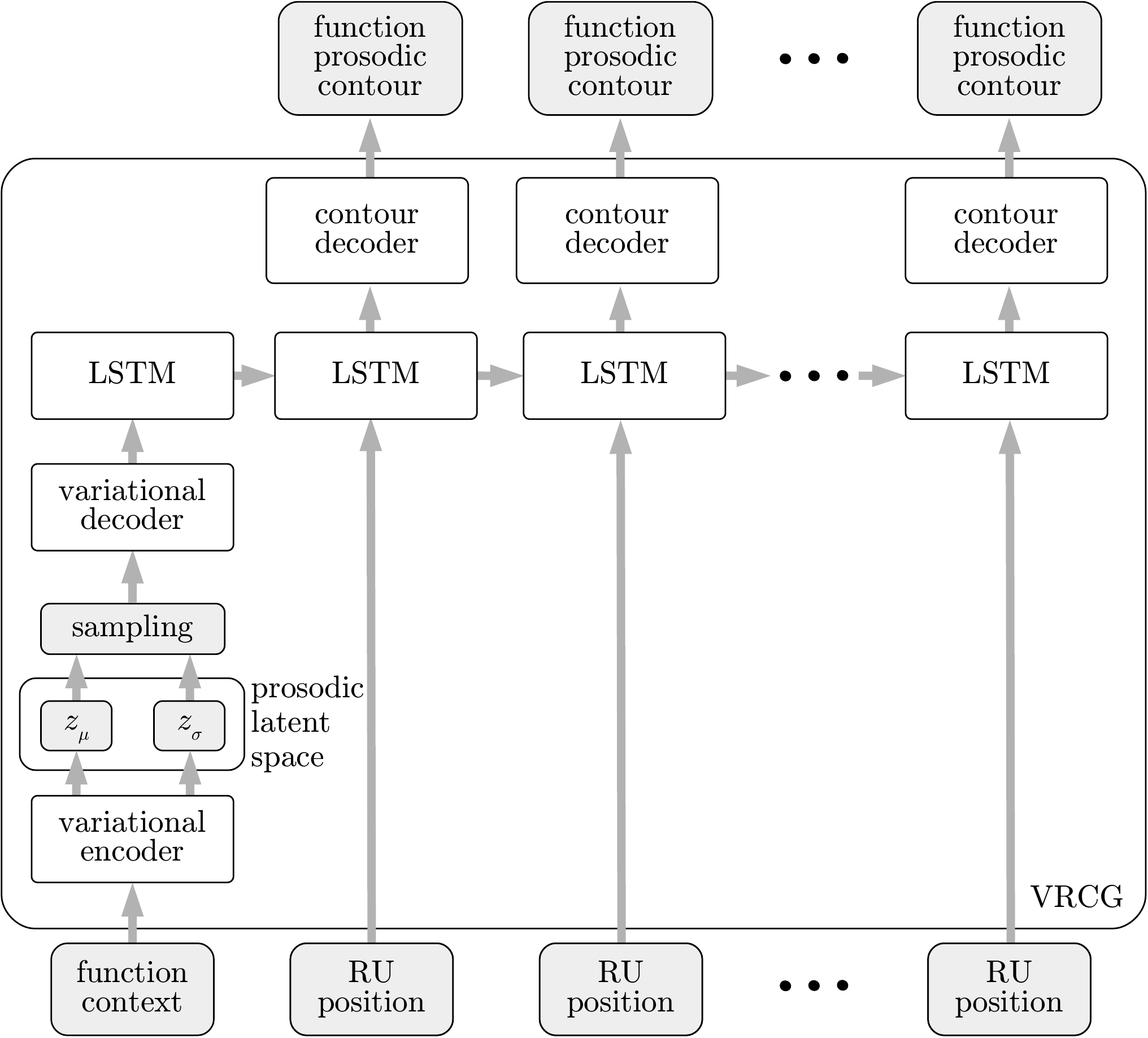}
 \caption{Comparison of the contour generator architectures (left to right): the WSFC weighted contour generator (WCG), and the VPM RU-based variational contour generator (VCG) and the variational recurrent contour generator (VRCG).}
  \label{fig:nncgs}
%     \end{minipage}
%     \hfill
\end{figure*}

The Weighted SFC model~\citep{gerazov2018wsfc}, introduces a weighted contour generator (WCG) that incorporates a weight module responsible for scaling its contribution in the decomposition, shown on the left in Fig.~\ref{fig:nncgs}.
This is similar to the Mixture of Experts (ME) model proposed by \cite{jacobs1991adaptive}, in that here the experts are the contour generators and the gates are the weighting modules.
The weight module in the WCG is itself a shallow neural network that computes the prosodic contribution of the contour given the context of its scope in the utterance.
Here, the context of the utterance can be arbitrarily defined, e.g. does it coincide with emphasis, or what is the attitude in the utterance.

By virtue of the WCG, the WSFC can model the prosodic prominence of the \emph{clich{\'e}s} in an utterance, such as the impact of the attitude and emphasis on the prominence of coinciding prototype contours~\citep{gerazov2018wsfc}.
This added degree of freedom also gives WSFC a slightly improved modelling performance.

In both the SFC and the WSFC, the CGs are trained within an analysis-by-synthesis loop.
In the synthesis part, the CGs are used to generate the prototype contours, which are then summed to form the utterance's prosodic contour reconstruction. In the analysis part, the reconstruction is subtracted from the original prosodic contour, and the error is distributed among the constituent CGs by adding it to their previous outputs.
These adjusted contours are used as new targets for training the CGs with backpropagation.
The training loop (synthesis, error distribution and CGs training) is typically iterated a dozen times.
Consistency of prototype contours obtained at convergence mainly depends on the statistical coverage of maximally independent locations and sizes of overlapping functions and scopes.
In the SFC the error is distributed equally to all the CGs, while in the WSFC their contribution, through their weight coefficients, is taken into account.

\section{The Variational Prosody Model}

The Variational Prosody Model (VPM) follows the SFC modelling paradigm but introduces two new features: \emph{i}) it introduces variational encoding to map a prosodic latent space able to model the context-sensitive variations of the prosodic prototypes, surpassing the one-dimensional modelling of prominence in the WSFC. Moreover, \emph{ii}) it integrates all the contour generators within a single network architecture. This allows the joint training of all of the contour generators, thus eliminating the need of an analysis-by-synthesis loop with its ad hoc distribution of errors.

\subsection{Variational Autoencoders}

An autoencoder (AE) is a deep neural network built to learn an efficient data encoding scheme, primarily used for data reduction \citep{cottrell1988principal, deng2010binary}, but can also be used for denoising \citep{lu2013speech}
and learning meaningful signal representations \citep{socher2011semi, liou2014autoencoder, ap2014autoencoder, obin2018sparse}.
The generic AE architecture comprises: an \emph{encoder} that receives the input signal and transforms it through a bottleneck layer to a latent low-dimensional representation, also called the latent code, and a \emph{decoder} which regenerates the input signal from the latent representation, giving the autoencoder its name.
The encoder and decoder are usually built from feedforward neural networks, but sequence-to-sequence models use recurrent neural networks that allow for variable length input/output signals, e.g. in machine translation \citep{cho2014learning, sutskever2014sequence}.

Variational autoencoders (VAEs), whose architecture resembles the variational contour generator (VCG) architecture of the VPM shown in the centre of Fig.~\ref{fig:nncgs}, differ from classical AEs in that their latent representation is probabilistic and thus continuous, allowing for random sampling and interpolation \citep{kingma2013auto}.
A variational encoder maps an input vector~$\mathbf{x}$ into a latent space representation $\mathbf{z}$ using an encoder neural network with parameters $\phi$ that outputs $q_\phi(\mathbf{z|x})$, i.e. a probability distribution of the hidden representation conditioned on the input.
In fact, $q_\phi(\mathbf{z|x})$ is an approximation of the intractable true posterior $p_\theta(\mathbf{z|x})$, which we assume takes a multivariate Gaussian form with a diagonal covariance matrix, i.e. for a given input data point $\mathbf{x}^{(i)}$:
\vspace{-5pt}
\begin{equation}
  q_\phi(\mathbf{z|x}^{(i)}) = \mathcal{N}(\mathbf{z};\bm{\mu}^{(i)}, \bm{\sigma}^{2(i)} \mathbf{I})
\end{equation}

Thus the output of the encoder network, for a given input $\mathbf{x}$ is a vector of $N$ means and $N$ variances, where $N$ is the chosen dimension of the latent space representation $\mathbf{z}$.
We can then sample the posterior distribution using the reparametrisation trick:
\begin{equation}
  \mathbf{z}^{(i,l)} = \bm{\mu}^{(i)} + \bm{\sigma}^{(i)} \circ \bm{\epsilon}^{(l)} \,\, , \quad \text{where} \, \bm{\epsilon}^{(l)} \sim \mathcal{N}(\mathbf{0, I}).
\end{equation}

The obtained sample can then be passed through the decoder neural network with parameters $\theta$, which models $p_\theta(\mathbf{x|z})$, and outputs an approximation of the original input vector $\mathbf{x}$. In fact, the parameters of the encoder and decoder networks $\phi$ and $\theta$ are trained using backpropagation and gradient descend so that the VAE reproduces as close as possible its input.
As a by-product of this process, the VAE learns the $q_\phi(\mathbf{z|x})$, structuring the latent space representation.

\subsection{Variational Contour Generators}

We explored two types of contour generators for the VPM: the RU\nobreakdash-based variational contour generator (VCG) and the variational recurrent contour generator (VRCG), both shown in Fig.~\ref{fig:nncgs}.
The variational encoding introduced in the contour generators of the VPM is reminiscent of the one used in VAEs.
Unlike the classic VAE architecture though, we do not use the output prosodic contour as input; thus they are not autoencoders per se.
Instead, we input the RU position and the function's context and train the contour generators to learn a context-specific latent space representation of the shape of each prosodic contour.
In the recurrent contour generator, which can be built with long short-term memory (LSTM) cells, the prosodic space mapping is additionally decoupled from the RU position within the scope, and only depends on the function's context.

\subsection{VPM architecture}

\begin{figure}[t]
  \setlength\belowcaptionskip{-10pt}
    \centering
%     \begin{minipage}[t]{.48\linewidth}
	\includegraphics[width=.88\columnwidth]{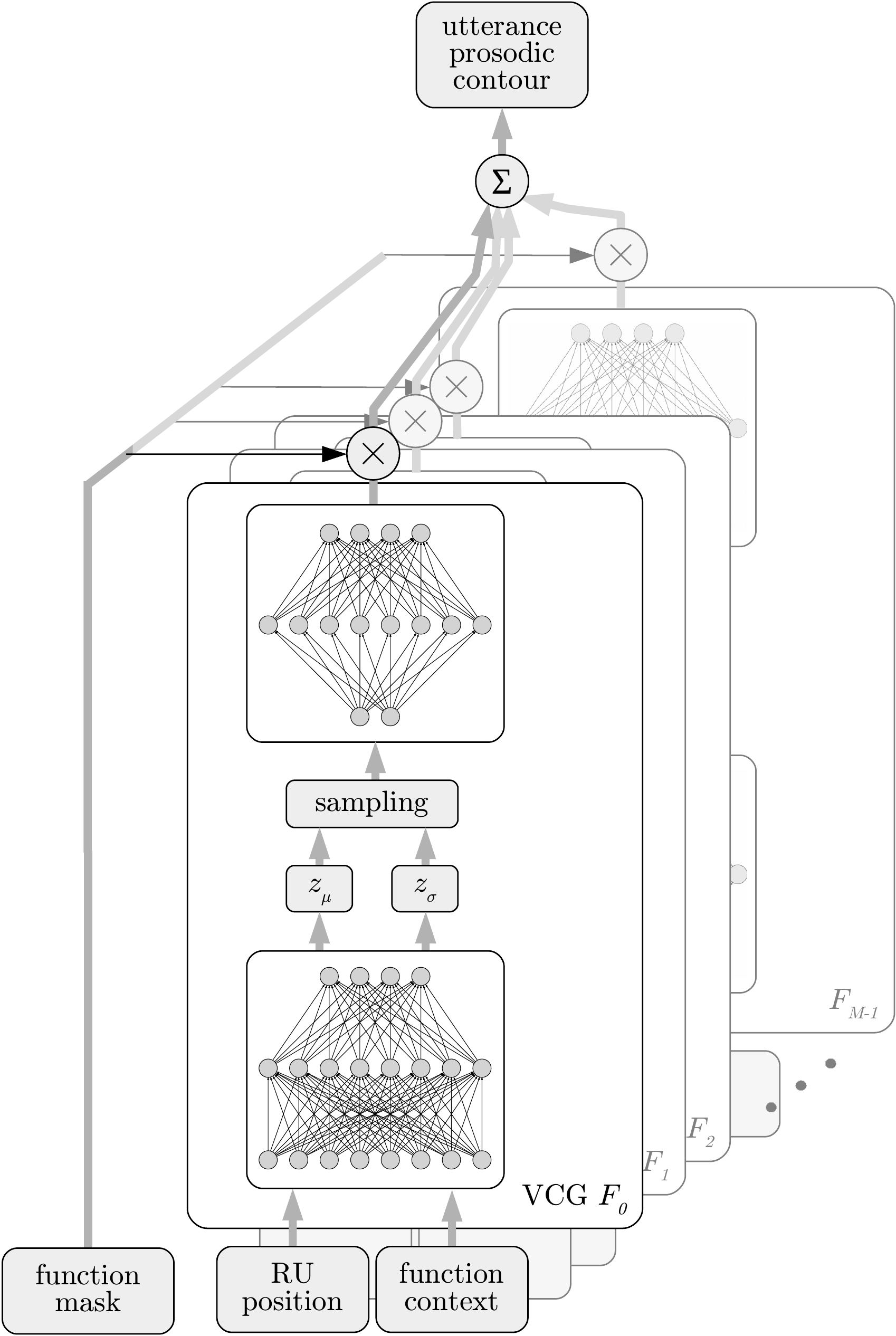}
      \caption{Static form of the Variational Prosody Model architecture, shown here built with RU\nobreakdash-based VCGs. The function mask is used at training time to solely activate the CGs that contribute to a given RU position. Contrary to the original SFC, the repartition of the modelling error is performed by gradient descent.}
	\label{fig:arch}
%     \end{minipage}
\end{figure}

The architecture of the VPM is intrinsically dynamic in that the combination of contour generators used depends on the linguistic functions in the utterance, i.e. for a given utterance, the VPM recruits the contributing contour generators and overlaps and adds their outputs accordingly.
A static structure of the VPM architecture can be imposed though, shown in Fig.~\ref{fig:arch}.
The static architecture improves training in two ways: \emph{i}) it gives the freedom to choose an arbitrary batch size, i.e. the number of samples of data used for a single update of the VPM parameters, one that is independent of the number of occurrences of a particular linguistic function combination.
This means that having a small number of occurrences of a particular combination will not result in a smaller batch size, producing an imbalance of the influence on parameter updates among the combinations.
And \emph{ii}) the shared graph structure across combinations allows the batches to contain a random set of different contour combinations, resulting in a better parameter update per batch.

The static architecture comprises copies of the contour generators necessary to cover all possible overlapping function combinations found within a dataset.
A mask is then used for each RU to take into account only the active contour generators, both for contour prediction in the forward pass, and for weight updates in the backpropagation.

\subsection{VPM training}

The loss function used to train the VPM comprises the Mean Square Error (MSE) of the utterance's prosodic contour reconstruction and the Maximum Mean Discrepancy (MMD) term for regularisation, as proposed in InfoVAEs~\citep{zhao2017infovae}:
\begin{align}
\begin{split}
  % \vspace{-40pt}
 \mathcal{L} &= \mathcal{L}_\text{MSE} + \lambda \mathcal{L}_\text{MMD} \\
 &= \frac{1}{N} \sum_{n=0}^{N-1} (f - \hat{f})^2 +
 \lambda D_{\text{MMD}}(q_\phi(\mathbf{z}) || p(\mathbf{z})) \, .
 \end{split}
 \label{eq:loss}
\end{align}
\vspace{-5pt}

\begin{figure*}[!t]
  \setlength\belowcaptionskip{-10pt}
 \centering
 \hfill
  \includegraphics[width=.3\linewidth]{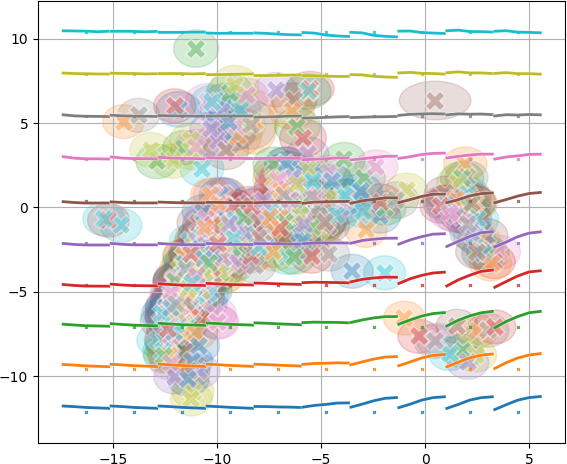}
  \hfill
  \includegraphics[width=.3\linewidth]{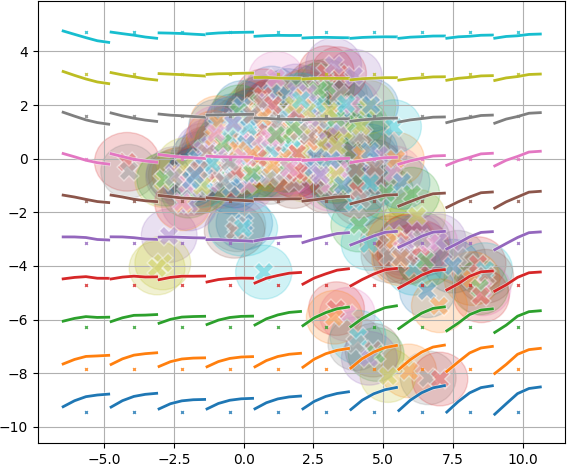}
  \hfill
  \includegraphics[width=.3\linewidth]{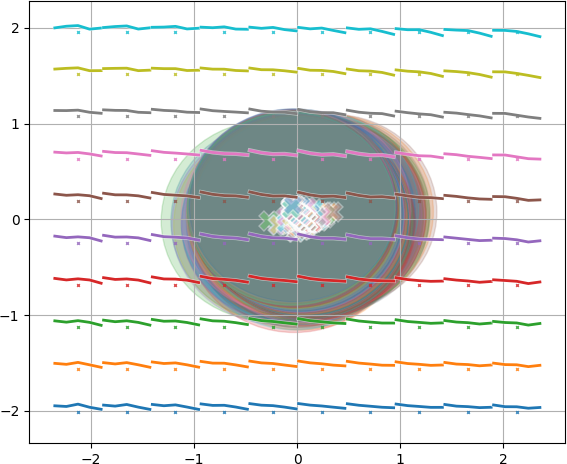}
  \hfill
  \hspace{0pt}

 \caption{Structure of the prosodic latent space for the left-dependency function contour (DG) from \small{\tx{Morlec}} obtained with a VPM using RU-based VCGs, for different values of the regularisation coefficient $\lambda$ (left to right): 0.01, 1, and 100.
 The X-s and the ellipses show the means and standard deviations of the latent space distributions corresponding to each unique input combination. The plots also show the contours generated when sampling the prosodic latent space.
 These graphs show that $\lambda$ controls the compactness of the latent representation}
\label{fig:reg}
\end{figure*}

Here, $f$ and $\hat{f}$ are the original and reconstructed prosody contours and $N$ is the number of prosody samples per rhythmic unit.
In our case we use $N-1$ pitch targets and a duration coefficient. Specifically, we use $N=6$, i.e. 5 $f_0$ pitch targets per vocalic nucleus, evenly distributed throughout its duration.
$D_{\text{MMD}}$ is the Maximum-Mean Discrepancy (MMD) divergence measure between the marginal inference distribution on the latent space $q_\phi(\mathbf{z})$ and the prior $p(\mathbf{z})$, and $\lambda$ is the regularisation coefficient.
The MMD quantifies the distance between two distributions by comparing all their moments, and can be calculated using the kernel trick:
\begin{align*}
 D_{\text{MMD}}(p||q) = & \,\, \mathbb{E}_{p(\mathbf{z}),p(\mathbf{z}')}[k(\mathbf{z},\mathbf{z}')]
 - 2\, \mathbb{E}_{q(\mathbf{z}),p(\mathbf{z}')}[k(\mathbf{z},\mathbf{z}')] \\
 &+\mathbb{E}_{q(\mathbf{z}),q(\mathbf{z}')}[k(\mathbf{z},\mathbf{z}')] \, . \numberthis \label{eq:mmd}
\end{align*}
We have used the MMD for regularisation instead of the Kullback–Leibler divergence (KLD) as proposed in the original Evidence Lower Bound (ELBO) criterion~\citep{kingma2013auto}, because it gave poor training results for the VPM.
In fact, it has been established that the KLD on the posterior $q_\phi(\mathbf{z|x})$, with a powerful enough decoder, leads the VAE to ignore the latent code, resulting in a latent space identical to the prior $p(\mathbf{z})$~\citep{chen2016variational}.
This issue in VAE training cannot be mitigated by simply scaling the KLD term as suggested in $\beta$-VAE \citep{higgins2016beta}, and has been addressed in various ways in literature: by limiting the power of the decoder~\citep{chen2016variational}, regularisation scheduling strategies for the KLD term~\citep{bowman2015generating}, and the use of divergence measures on the $q_\phi(\mathbf{z})$ as in InfoVAEs~\citep{zhao2017infovae}.

\section{Experiments and results}
We have designed our experiments to test five hypotheses:
\emph{i}) the VPM is a plausible model that can decompose the prosodic contour into its constituent prototype contours,
\emph{ii}) the VPM network architecture, because of its joint contour generator optimisation, is able to outperform the analysis-by-synthesis based SFC and WSFC when using the same contour generator structure, while maintaining the same level of performance with a standard deep model,
\emph{iii}) the VPM is able to capture contour prominence as well as the WSFC,
\emph{iv}) the VPM, unlike the SFC and WSFC, is able to capture context-specific variance in the shape of the extracted prototype contours, and
\emph{v}) the VPM is also able to capture variance in the shape of the extracted prototype contours that is not context-specific.

All of the models used in these experiments have been implemented in the scientific Python ecosystem using the PyTorch deep learning library \citep{paszke2017automatic} and the Scikit-Learn machine learning library \citep{scikit-learn}. The code is available as free software on GitHub.\footnote{\url{https://github.com/gerazov/prosodeep}}

\subsection{Databases}

To test the four hypotheses, we use three databases in our experiments:
\vspace{-3pt}
\begin{itemize}
\setlength\itemsep{0em}
\item \tx{Morlec} -- a database of 6 attitudes in French: declaration, question, exclamation, incredulous question, suspicious irony and obviousness, totalling 1932 utterances from one speaker~\citep{morlec2001generating},
\item \tx{Liu} -- a database of declarations and five question types
in Chinese that include emphasis at three different positions \citep{liu2005parallel}. We use the first speaker with 76 carrier sentences using a single tone each,
recorded 5 times, totalling 380 utterances, and
 \item \tx{Chen} -- a database of read Chinese from a single female speaker comprising 110 carrier utterances ranging from 6 to 38 syllables in length~\citep{chen2004superposed}.
\end{itemize}

\subsection{Hyperparameters}

The most important hyperparameters in training the VPM are the dimension of the latent space, the number of hidden layers and their size, and the regularisation coefficient $\lambda$ in the loss function~\eqref{eq:loss}.
In our experiments we use a two-dimensional prosodic latent space, both because it offers ample modelling power and because it is favourable for visualisation and exploration.
The number of hidden layers and their size also impacts the modelling power of the system, if it is too small the model will have trouble handling the data complexity and if it is too high the model will have a tendency to overfit the training data.
Also, the larger the number of layers and the larger their size, the more data the model needs to train all of its parameters.

\begin{figure*}[t]
  \setlength\belowcaptionskip{-10pt}
  \centering
  \includegraphics[width=.325\linewidth]{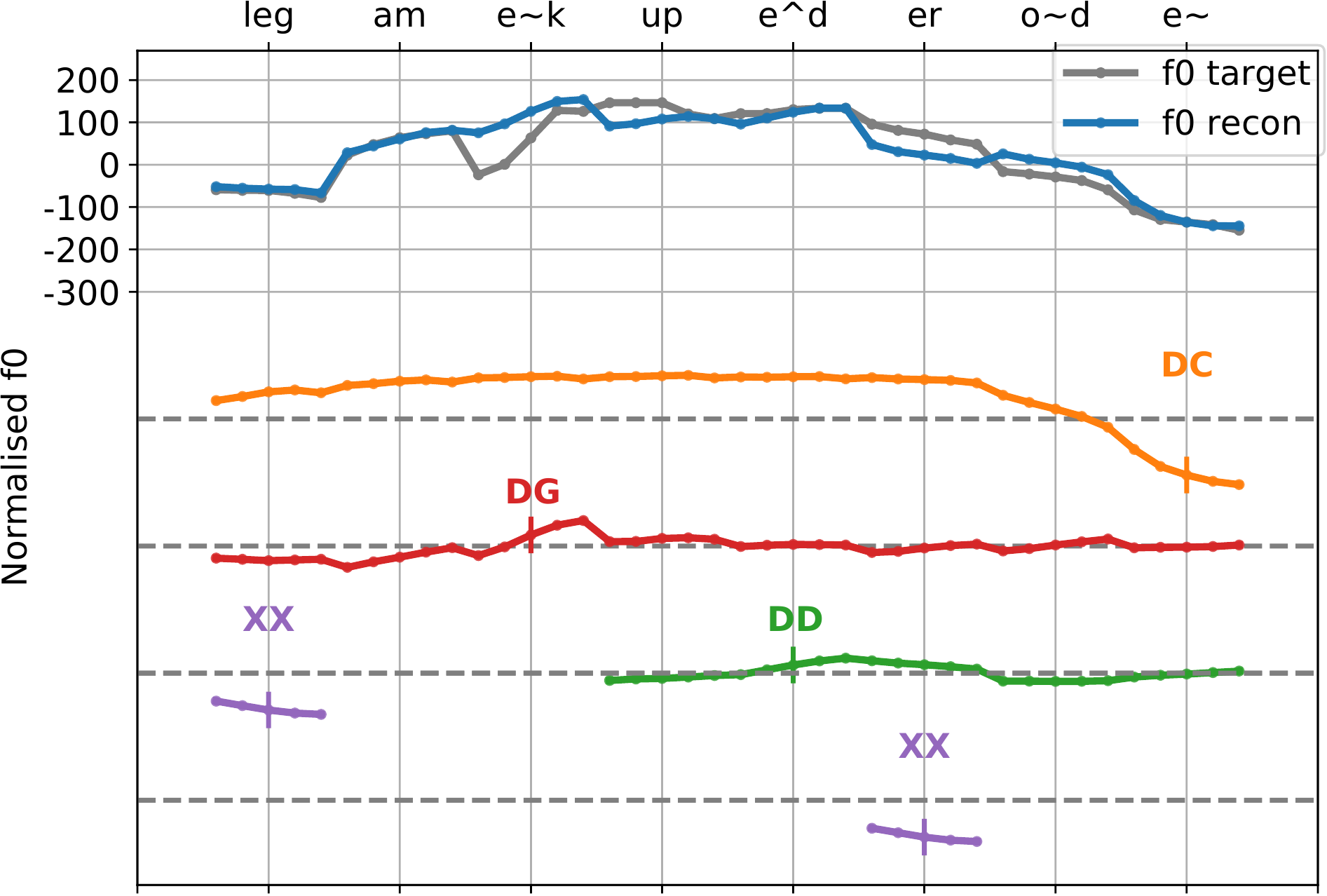}
  \hfill
  \includegraphics[width=.325\linewidth]{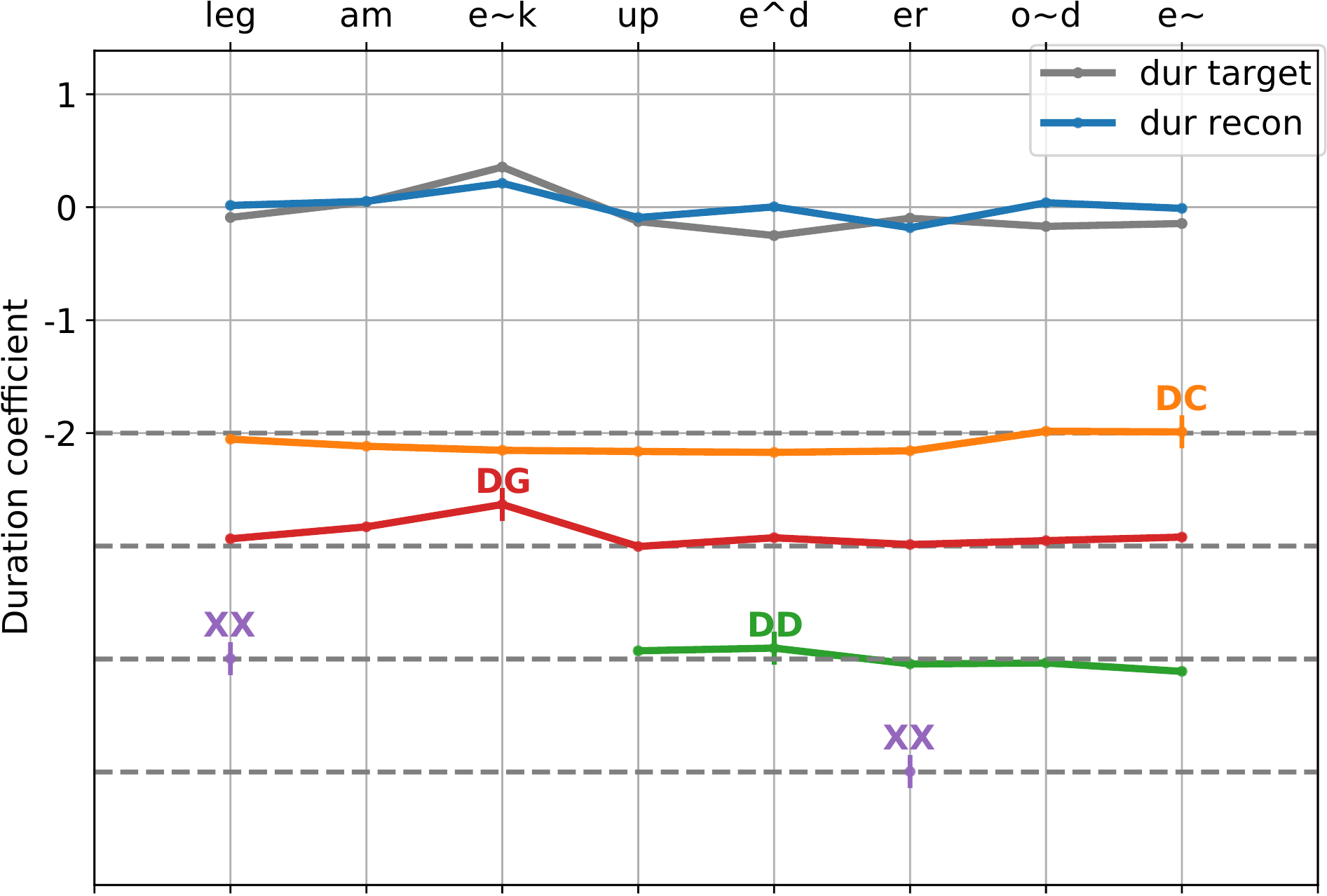}
  \hfill
  \includegraphics[width=.325\linewidth]{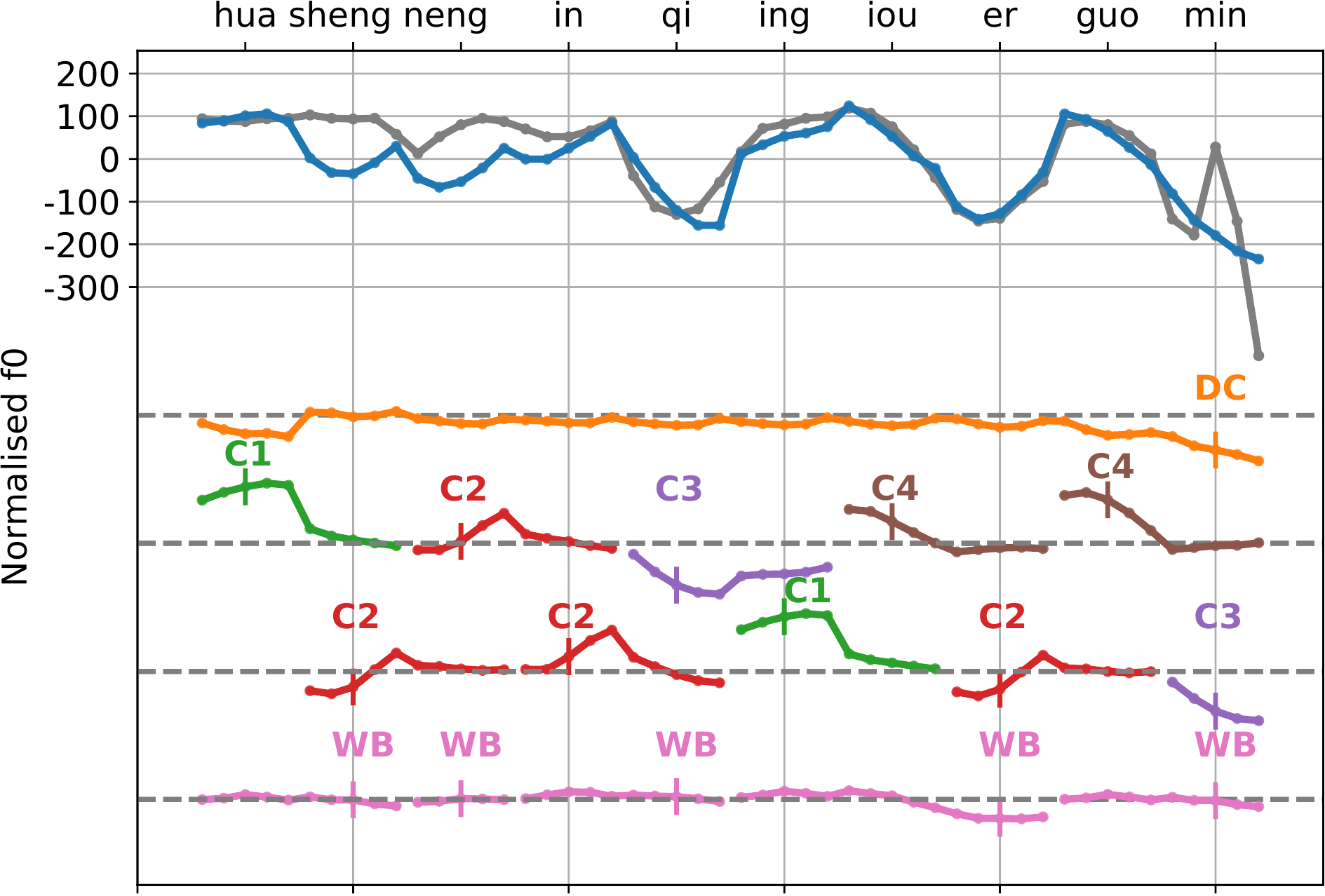}

  \includegraphics[width=.325\linewidth]{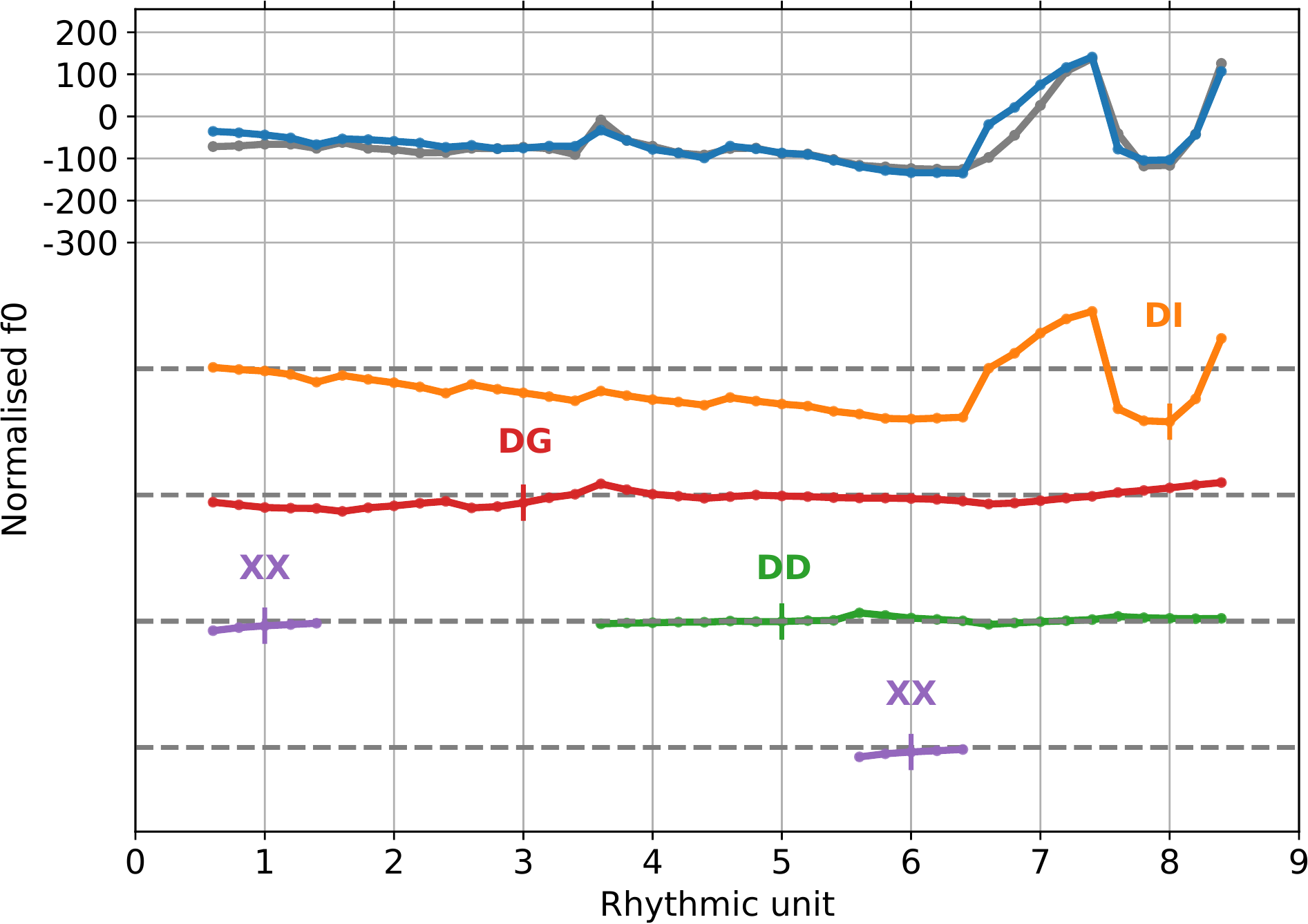}
  \hfill
  \includegraphics[width=.325\linewidth]{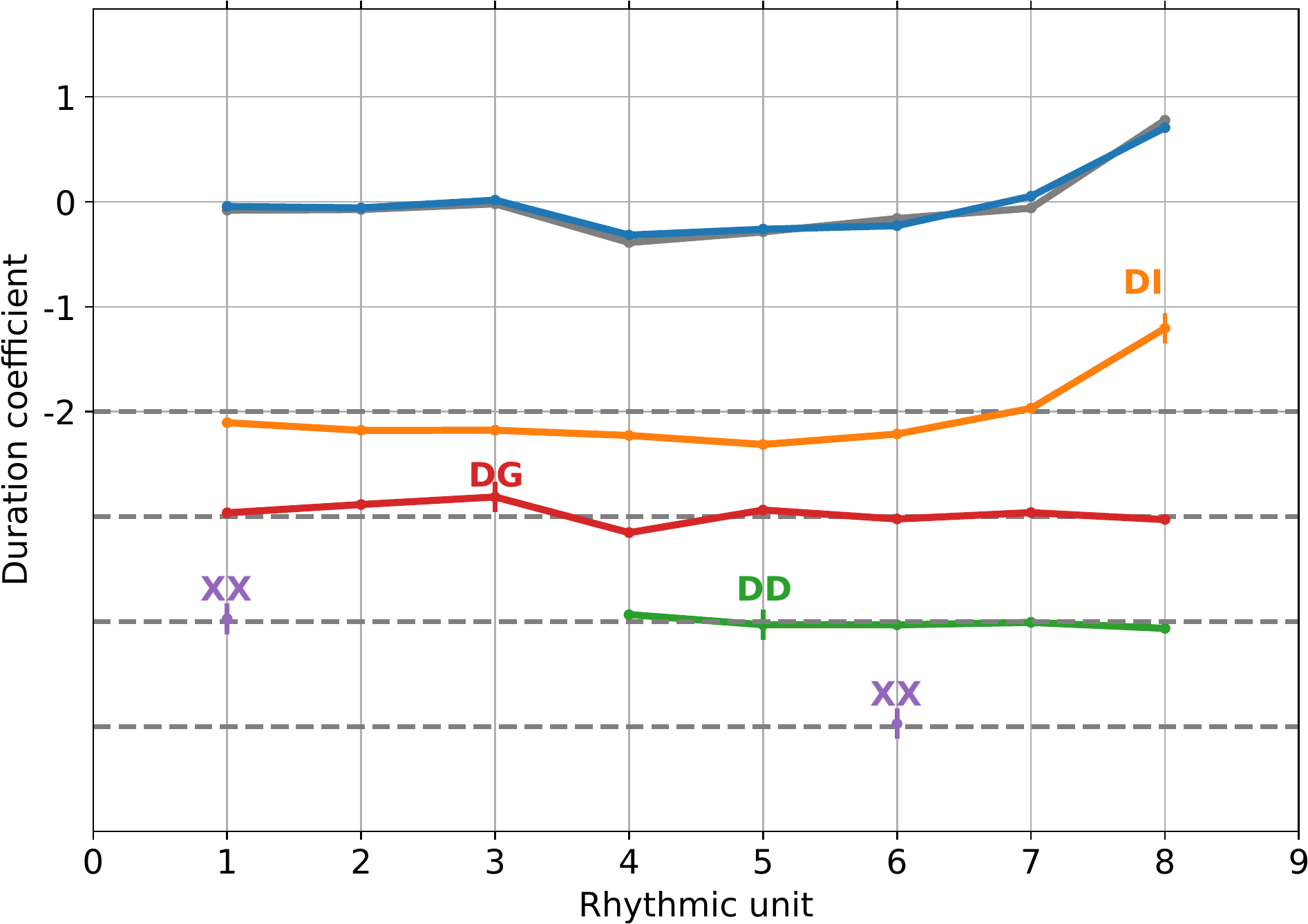}
  \hfill
  \includegraphics[width=.325\linewidth]{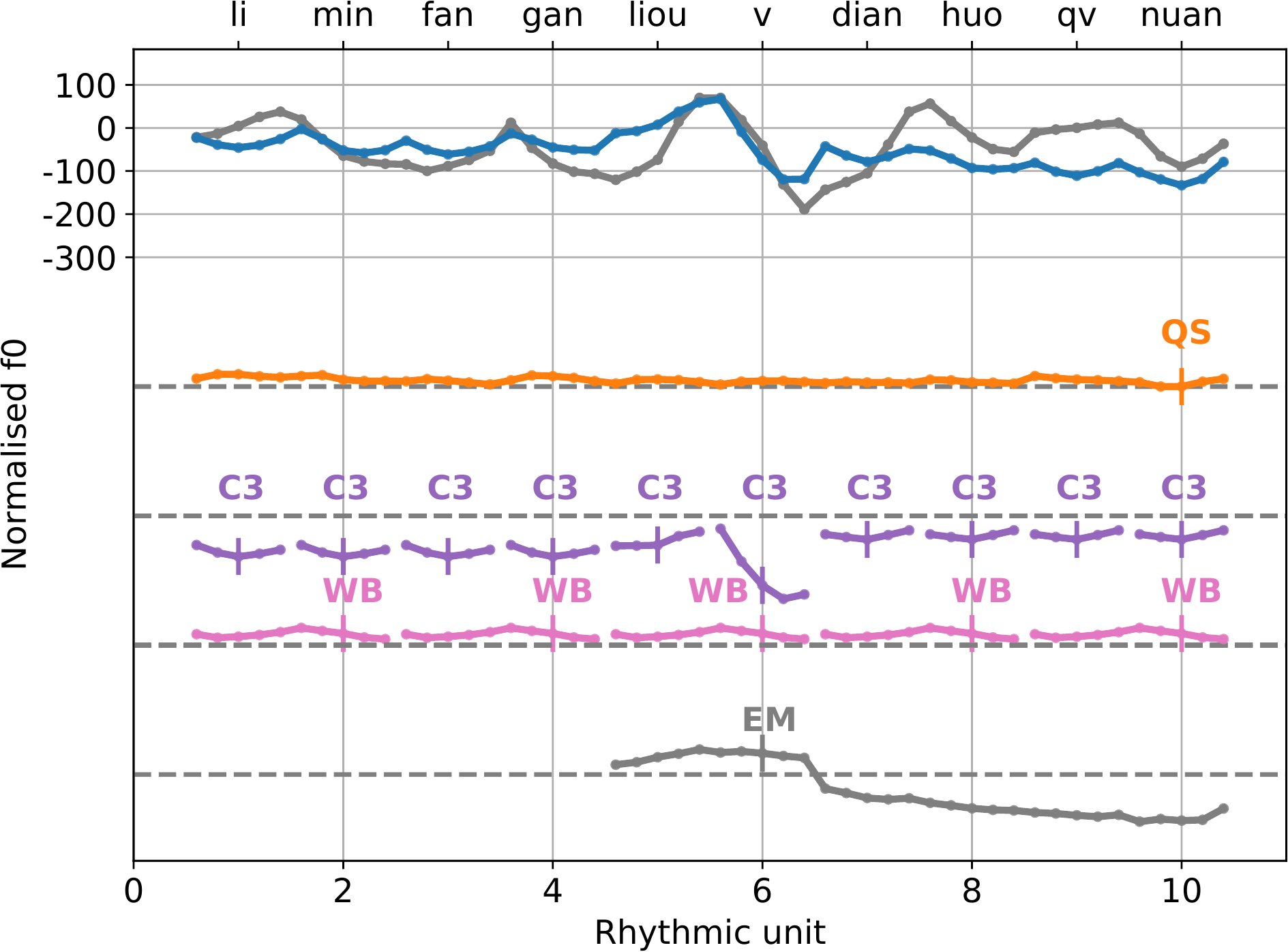}
  \caption{Decomposition of the melody (left column) and RU duration modification coefficient (centre column) of the French utterance
  \emph{``Les gamins coupaient des rondins.''} from \small{\tx{Morlec}}, with the VPM\nobreakdash-VRCG for declaration (DC, top row) and incredulous question (DI, bottom row) into constituent prototype contours: attitude (DC and DI), left- and right- dependencies (DG, DD), and clitics (XX).
  The VPM successfully captures the strongly reduced prominence of XX, DG and DD, when solicited in the DI context.
  Decomposition of the intonation of two Chinese utterances (right column):
  \emph{``Huā shéng néng ín qǐ īng iòu ér guò mǐn.''} from \small{\tx{Chen}} (top) and \emph{``Lǐ Mǐn fǎn gǎn {\bf \emph{Liǔ Yǔ}} diǎn huǒ qǔ nuǎn?''} from \small{\tx{Liu}} (bottom)
  with the VPM\nobreakdash-VRCG, a declaration (DC) and a question (QS) with the four Chinese tones (C1 - C4), word boundaries (WB), and emphasis (EM). The VPM captures the correct shapes of the tones and properly captures the effect emphasis has on them.}
  \label{fig:example}
\end{figure*}

The regularisation coefficient can compact the prosodic latent space, but it can also hinder learning meaningful contours by forcing the mapped distributions to overlap.
This process is illustrated in Fig.~\ref{fig:reg}, which shows the changes in the prosodic latent space mappings of the unique input combinations for the left-dependency functional contour (DG) in \tx{Morlec} as a function of the regularisation coefficient for a VPM with RU\nobreakdash-unit based VCGs.
We can clearly see the convergence of the mapped distributions in the latent space as $\lambda$ increases, both from their relative dispersal, as well as the scale of the latent space axes.
Also, a $\lambda$ of 100 can be seen to collapse the latent space structure into the prior, thus precluding learning of a latent space that would adequately represent the variability of the elementary contours.

To obtain the best value for these as well as other hyperparameters we conducted a grid-search over the \tx{Morlec} data, as it has the sufficient prosodic variety that we want our model to handle.
We evaluated the model's performance using a validation set and found that we obtained best performance for a single hidden layer with a size of 32 units, a $\lambda$ value of 0.3, $\tanh$ activation function, a $L^2$ norm weight decay of 0.0001, a batch size of 256 RUs, and a learning rate of 0.001 with the Adam optimizer \citep{kingma2014adam}.
For the recurrent contour generators we chose a batch size of 8 utterances. The RU position ramps were not normalised because we found that normalisation in the range 0.01~--~0.99 \citep{wu2016merlin}, did not lead to performance gains for both \tx{Morlec}, as well as \tx{Chen}, which contains longer utterances.
These hyperparameter values were used in our further experiments.

We empirically verified that LSTM cells work better than gated recurrent units (GRUs) and vanilla recurrent neural networks (RNNs). In addition, we determined that the best way to bias the LSTM with the latent space sample is to feed the sampled value as an initial input step, similar to image captioning applications \citep{vinyals2015show} and as shown in Fig.~\ref{fig:nncgs}.

\subsection{Baseline}

The baseline system used for comparison was built using the benchmark intonation module in the Merlin speech synthesis system \citep{wu2016merlin}.
We used the basic deep neural network (DNN) model and the single-direction recurrent model based on LSTM cells. We chose the latter as it outperformed the bi-directional LSTM models for modelling intonation in their benchmark.
In the original Merlin implementation the DNN comprises 6 feedforward hidden layers, 1024 units each, while the LSTM model comprises five feedforward hidden layers of 1024 each, with 512 units for the last one. Their output is based on some 416 features containing the answers to 416 binary and numerical questions about the context of the phone, all normalised to 0.01~--~0.99. In addition, there are 6 frame and state position features and 3 state and phone duration features.

To make a fair comparison with the VPM, in our implementation we feed the two Merlin models with the information about the type of overlapping functions at each RU, and the RU positions within the scope of each of these functions.
To avoid problems with the dataset size we ran a grid search on the baseline model hyperparameters and chose the best performing configuration on the \tx{Merlin} dataset.
We found that for the DNN model, two hidden layers of 256 units each, gave the best results.
For the LSTM model it was one hidden layer with 1024 units.
The rest of the hyperparameters for the models were kept the same as the benchmark: $\tanh$ activation, $10^{-5}$ $L^2$ regularisation, and a learning rate of 0.002.

\subsection{Plausibility}

The plausibility of the proposed VPM can be qualitatively observed in the example decomposition of the two French utterances and two Chinese utterances in Fig.~\ref{fig:example}. The intonation decomposition of the example utterances is shown in the left and right columns, while the centre column plots the duration coefficients for the French utterances. In these decompositions, the latent prosodic space of the VPM was sampled at the means of the distributions.

\begin{figure}[t]
  \setlength\belowcaptionskip{-10pt}
  \centering
  \includegraphics[width=.9\columnwidth]{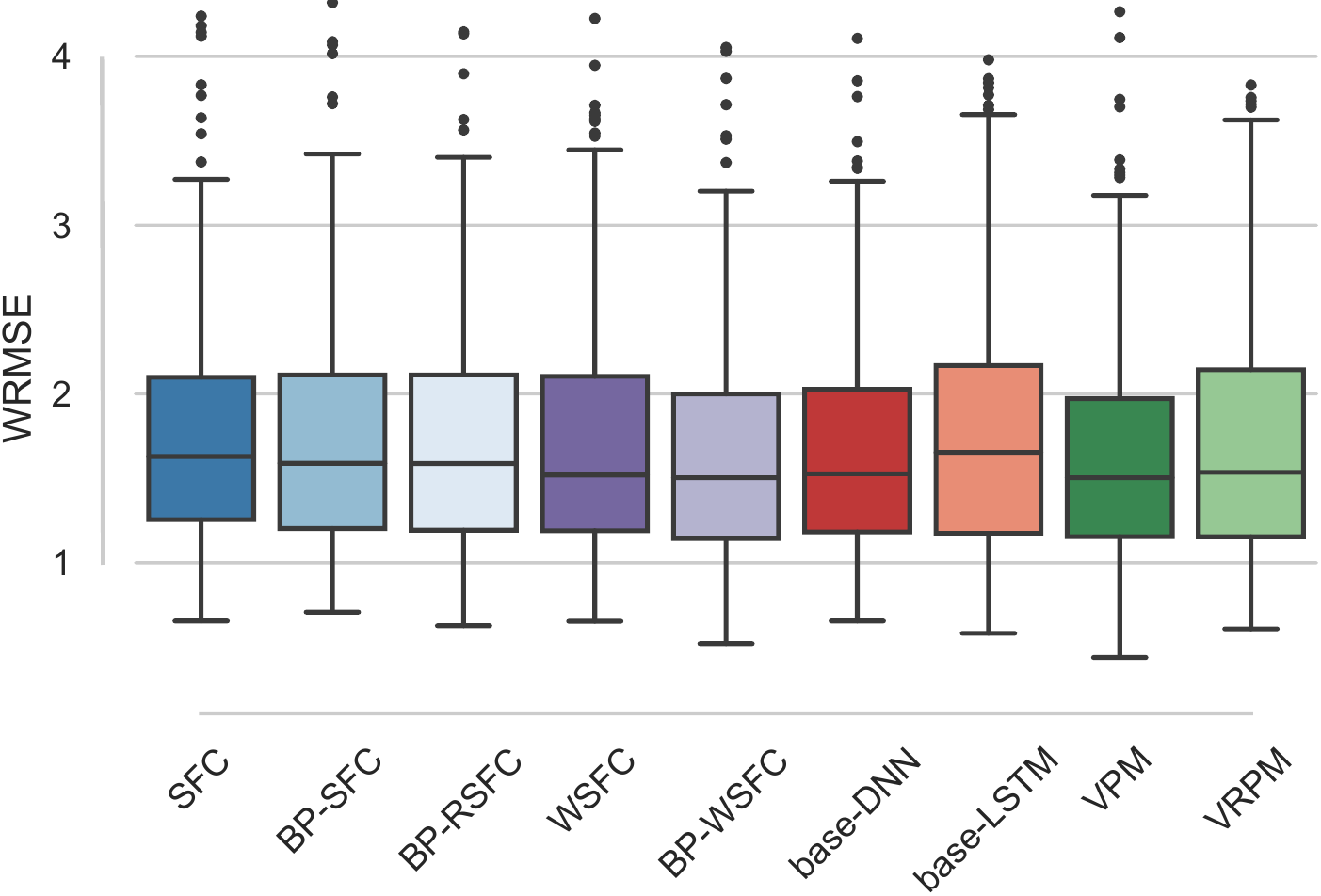}
  \vspace{10pt}

  \includegraphics[width=.9\columnwidth]{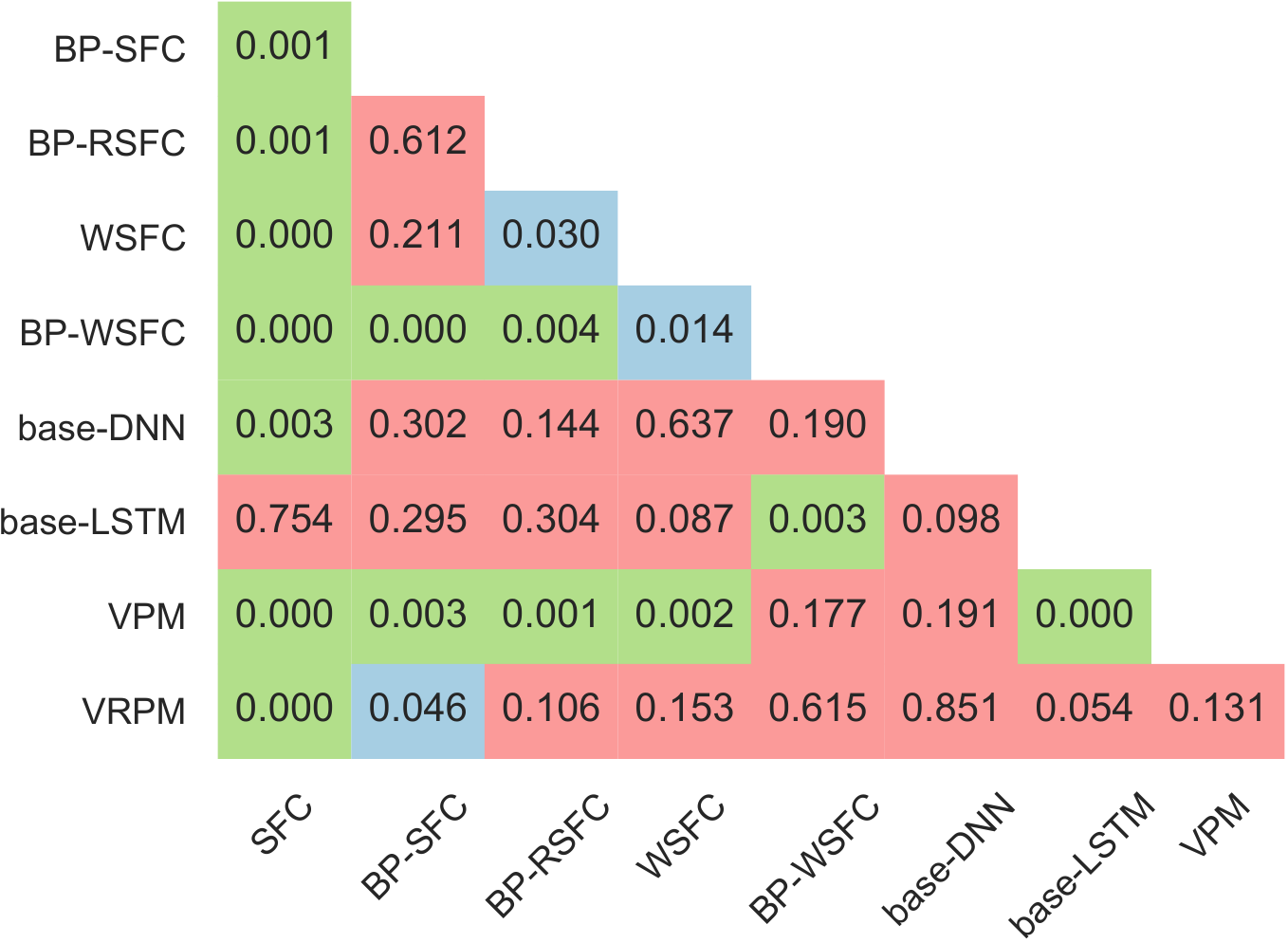}
  \caption{WRMSE distribution for the different models for \small{\tx{Morlec}} and the $p$-values for these distributions between each pair of models.}
  \label{fig:perf}
\end{figure}

We can see that the model successfully extracts the general shape of the prototype pitch contours from the intonation contour, specifically the shapes of the four Chinese tones. Also, we can see that the model correctly captures the central speed up and then the phrase final slow down of the speech rate for the incredulous question through the attitude contour itself, and not via the contributing syntactic contours.
Finally, and most importantly, we can see that the difference in prominence of the syntactic contours between the two attitudes in French, and the tone contours in context to the position of emphasis in the second Chinese example, is successfully captured by the VPM through its variance encoding mechanism. This is similar to the results obtained with the WSFC that has an explicit prominence weighting mechanism \citep{gerazov2018wsfc}.
All of this confirms our initial hypothesis that the VPM is a plausible prosody model able to decompose the prototype contours.

\begin{figure*}[t]
  % \begin{minipage}[t]{.4\textwidth}
  \setlength\belowcaptionskip{-10pt}
  \centering
    \hfill
  \includegraphics[height=5.8cm]{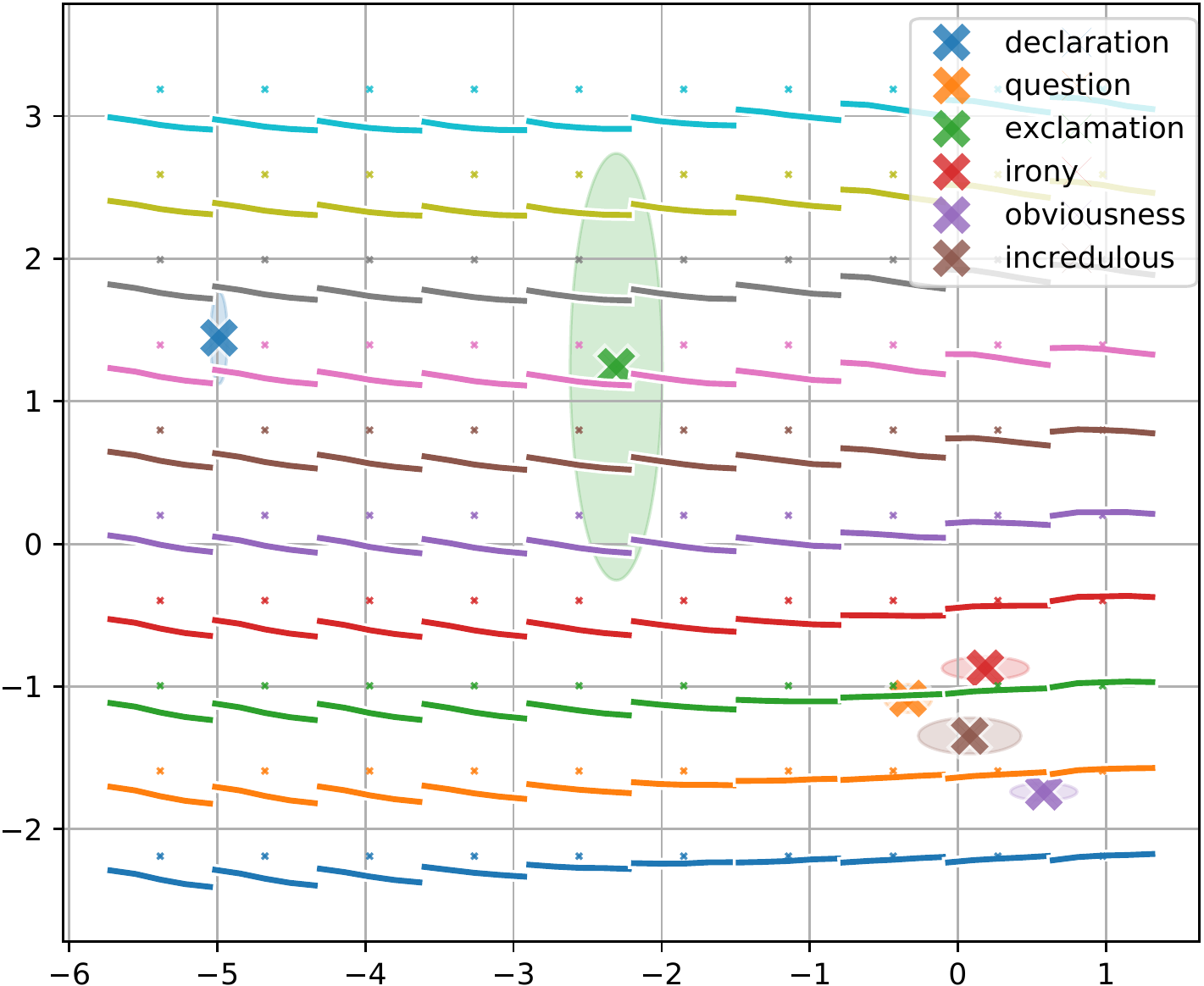}
  % \label{fig:xxcontours}
% \end{figure}
% \end{minipage}
% \hfill
% \begin{figure}[t]
% \begin{minipage}[t]{.57\textwidth}
  % \centering
  \hfill
  \includegraphics[height=5.8cm]{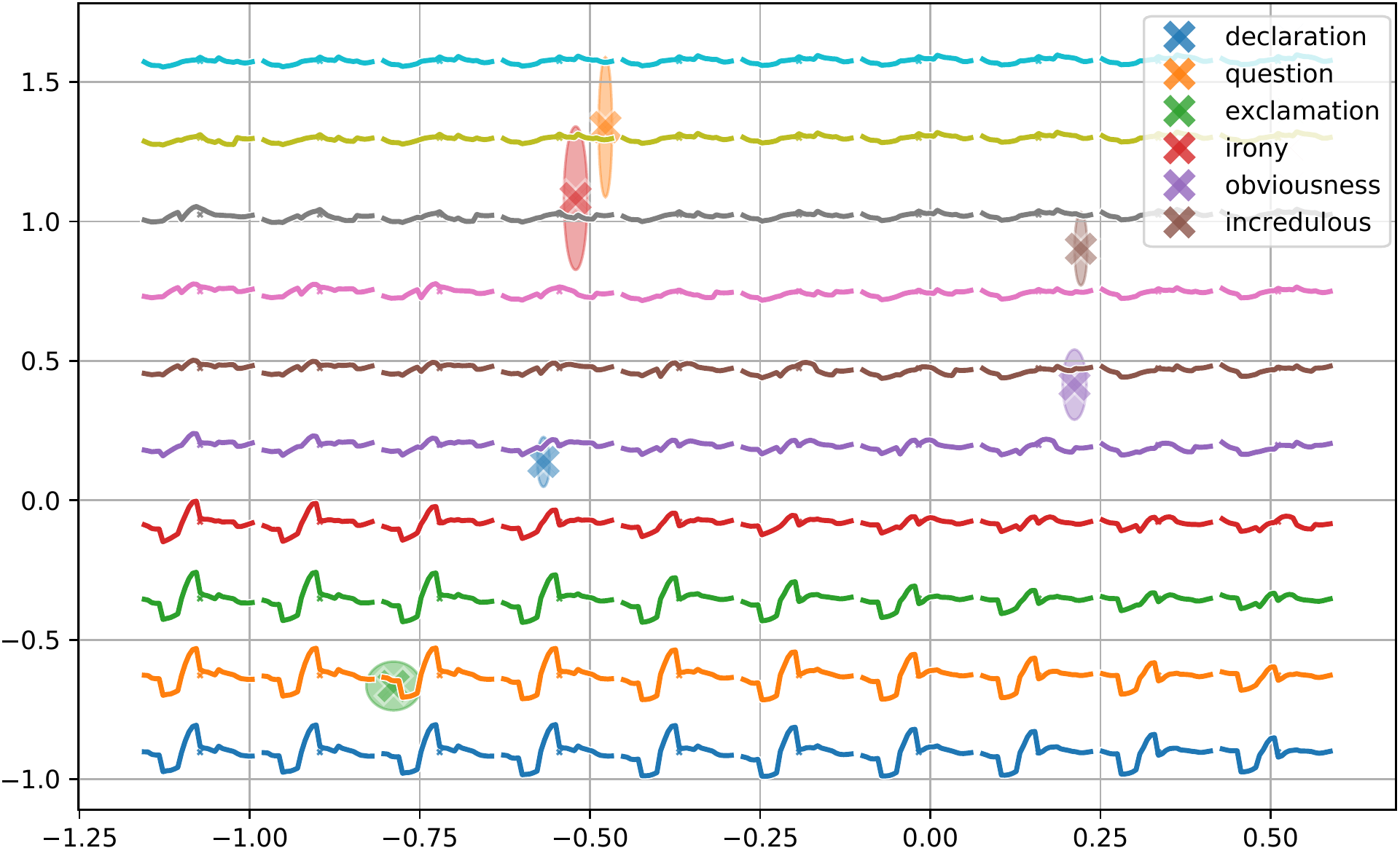}
    % \hfill % doesn't work at end of figure?
    \hspace{10pt}

  \caption{Prosodic latent space of the clitic functional contour (XX) in \small{\tx{Morlec}} captured by the VPM\nobreakdash-VCG (left) demonstrates that a full blown contour is only generated in context of the declaration and exclamation attitudes, while it is largely diminished for the rest: question, suspicious irony, obviousness, and incredulous question.
  % }\caption{
  Prosodic latent space of the left-dependency function contour (DG) in \small{\tx{Morlec}} captured by the VPM\nobreakdash-VRCG (right) shows that
  based on attitude context
   % with attitude codes same as in Fig.~\ref{fig:xxcontours}
   declaration and exclamation elicit full-blown contours, while they are largely diminished for the rest.}
  \label{fig:prominence}
% \end{minipage}
\end{figure*}

\subsection{Performance}
We assess the performance of the proposed VPM architecture by comparing its intonation reconstruction performance to that of the SFC and WSFC models, as well as the two baseline models on the \tx{Morlec}.
As a metric we have used a weighted version of the root mean square error (WRMSE) to take into account only the error in pitch within the nuclei.
To evaluate the impact on performance of the joint contour generator optimisation through backpropagation (BP), the VPM network architecture was also fitted with SFC contour generators (BP\nobreakdash-SFC), WSFC contour generators (BP\nobreakdash-WCG), as well as recurrent contour generators (BP\nobreakdash-RSFC).
The VPM in turn was evaluated both with the RU\nobreakdash-based VCGs (VPM), and the recurrent VRCGs (VRPM).

To maintain the overall model complexity we chose to keep the number of hidden units across the models to 32, except for the baseline models for which we kept the best performing parameters for the grid-search, i.e. two layers with 256 units for the base-DNN, and a single layer of 1024 units for the base-LSTM. Since there are 13 different contour types in the \tx{Morlec} data, the equivalent number of hidden units for the contour generator based models is around 416 units (32 hidden units $\times$ 13 contour generators). This number is indicative of the number of units that output the contour, but does not include the hidden units in the weight module of the WSFC, the encoder in the VCG, and the encoder and decoder in VRCG.
% Thus, the baseline models have a larger modelling capacity and with it a slight advantage over the other models.
All models were trained using early stopping based on a validation set, with a $\Delta$WRMSE threshold of $10^{-4}$ and a patience of 20 epochs, and then evaluated on a separate test set. The latent space was sampled at the mean of each mapping for this evaluation.

Table~\ref{tab:error} shows the obtained WRMSEs for the different models.
We can see that the VPM architecture when used with the SFC and WSFC contour generators outperforms the analysis-by-synthesis loop based training.
We can also see that the two varieties of the VPM generally outperform all of the other models.
The exception to this is that BP\nobreakdash-WSFC outperforms the VRPM.
We believe this is due to: \emph{i})~both the efficiency of capturing prominence by the weighted contour generators and prominence being the dominant prosodic variation in the prototypes in \tx{Morlec}, and \emph{ii})~the suboptimal extraction of the mean contours in the VPM induced by the random sampling procedure within training.
This effect is reduced when using the VCGs, where we have a spread of mappings for each RU position combination, as can be seen in Fig.~\ref{fig:reg}.
This more complex utilisation of the prosodic latent space, leads to larger modelling capacity, as is reflected in the better performance of the VCGs over the VRCGs.

\begin{table}[t]
  \centering
  \small
  \caption{Weighted RMSE in semitones for the reconstructed pitch obtained for \small{\tx{Morlec}} with the VPM compared to other models.}
  % \vspace{-10pt}
  %% local settings
  % \sisetup{detect-weight,mode=text}
  % for avoiding siunitx using bold extended
  \renewrobustcmd{\bfseries}{\fontseries{b}\selectfont}
  \renewrobustcmd{\boldmath}{}
  % abbreviation
  \newrobustcmd{\B}{\bfseries}
  % shorten the intercolumn spaces
  \addtolength{\tabcolsep}{-4.1pt}
  \begin{tabular}{l c}
    \multirow{2}{*}{Model}
    & {WRMSE} \\ & mean $\pm$ standard deviation \\
    \toprule
    \toprule
    SFC & $ 1.83 \pm 0.90$ \\
    BP-SFC & $1.76 \pm 0.85$ \\
    BP-RSFC & $1.77 \pm 0.81$ \\
    \midrule
    WSFC & $1.76 \pm 0.85$ \\
    BP-WSFC & $1.69 \pm 0.76$ \\
    \midrule
    base-DNN & $1.76 \pm 0.87$ \\
    base-LSTM & $1.83 \pm 0.94$ \\
    \midrule
    \B VPM & \B 1.68 $\pm$ 0.75 \\
    VRPM & $1.75 \pm 0.83$ \\
  \end{tabular}
  \label{tab:error}
  \vspace{-10pt}
\end{table}

The distribution of the WRMSE within the test set for each model is shown in the boxplot in Fig.~\ref{fig:perf}.
To analyse the statistical significance of the difference in results obtained with the different models, we used a post-hoc analysis of Wilcox’s Robust Repeated-Measures Analysis of Variance (RM-ANOVA) \citep{wilcox2011introduction}.
The obtained $p$-values are shown in the heatmap in Fig.~\ref{fig:perf}.
We can see that although for some models we get statistically significant differences in the performance result, e.g. for the SFC and the VPM, most of the results are not significantly different, e.g. for the VRPM.
This leads us to believe, that the models share a similar level of performance.
Even so, the presented findings confirm our second hypothesis that the joint contour generator training leads to performance benefits, and that the VPM performs on a par with, if not slightly better than, a standard deep baseline.

\subsection{Prominence}

The ability of the VPM to capture prominence due to attitude context in \tx{Morlec} has already been shown in the example decomposition shown in Fig.~\ref{fig:example}. Here, we reaffirm this observation by exploring the prosodic latent space of the clitic contour (XX) solicited in the 6 different attitudes, shown to the left in Fig.~\ref{fig:prominence}, and of the left-dependency contour (DG) shown to the right.
In both cases, we can clearly see that the declaration and exclamation attitudes map to areas in the latent space where the prototypes are fully realised.
On the other hand, all of the other attitudes map to areas where these contours are largely diminished and close to 0.
In fact, we can argue that there is a ``prominence'' vector in these latent spaces captured by the VPM.
Thus the VPM captures gradience as well as WSFC, but it is also able to model more subtle variations of shape.
Note that, since the DG prototype contour has a larger scope, we have to use the variational recurrent contour generators that decouple the latent space from the scope position ramps.

To analyse the impact of emphasis, i.e. narrow focus, on Chinese tones we will use the \tx{Liu} database.
Even though the carry-over effect has a significant impact on modelling performance~\citep{gerazov2018tones}, the imposed uniform structure of the utterances in \tx{Liu} precludes training tonal prototypes for an expanded scope. Thus we trained tonal prototypes with a single RU scope.
To demonstrate the usability of the VPM in a transfer learning scenario, we first pretrained the tonal prototype contour generators on the \tx{Chen} data, which contains a natural tone distributions,
and then fine-tuned them on the \tx{Liu} data.

\begin{figure}[t]
  \setlength\belowcaptionskip{-10pt}
  \centering
  \includegraphics[height=5.8cm]{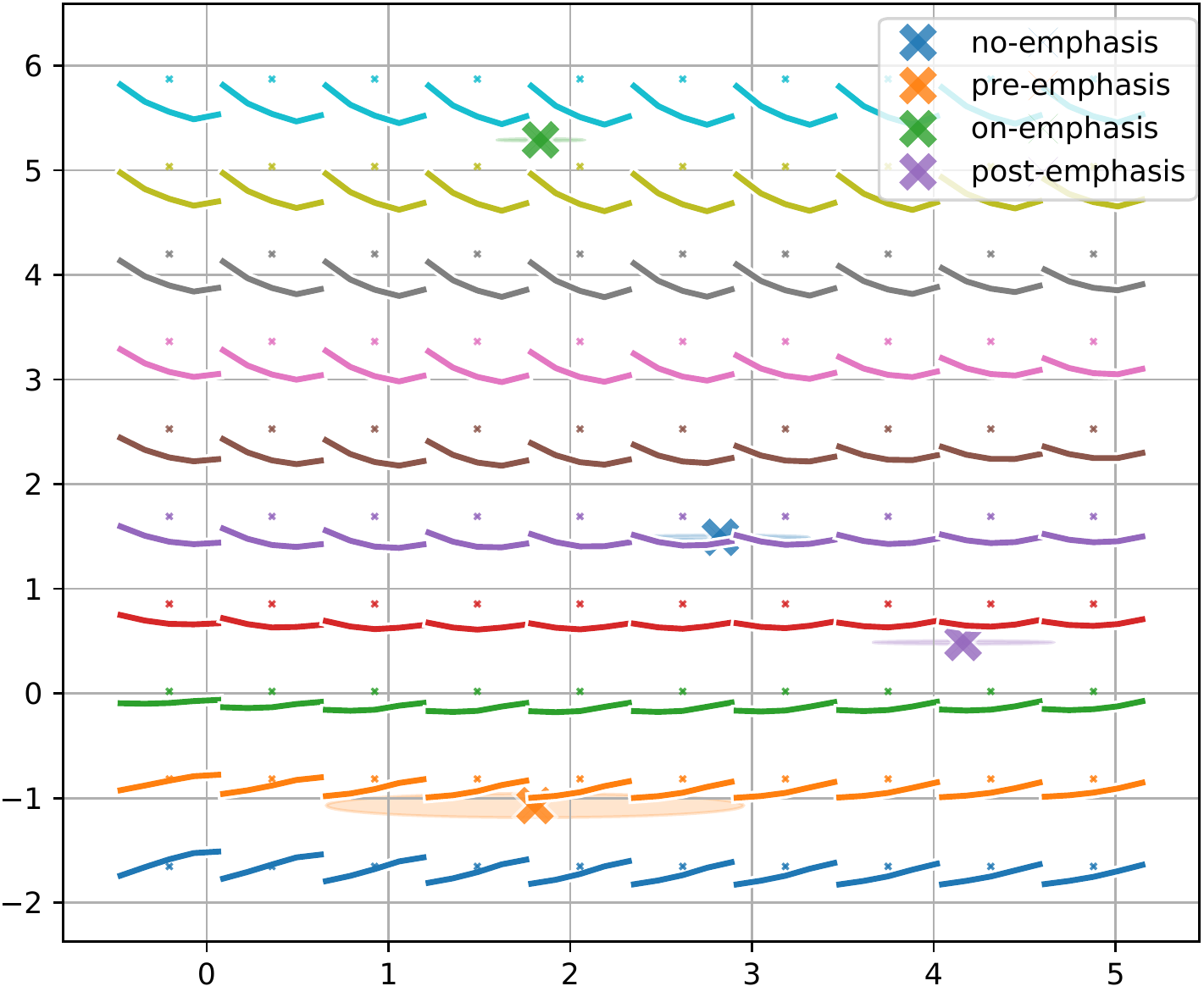}
  \caption{Prosodic latent space of the Tone 3 prototype contour in \small{\tx{Liu}}, in context of no-, pre-, on- and post-emphasis, as extracted by the VPM-VCG. The VPM captures prominence in the on-emphasis context, as well as shape variation in the pre- and post-emphasis contexts.}
  \label{fig:emph}
\end{figure}

The structure of the latent space for Tone 3 is shown in Fig.~\ref{fig:emph}. Note that this latent space corresponds to the decomposition shown in the bottom left plot of Fig.~\ref{fig:example}. Since the focus in the \tx{Liu} data always falls on two consecutive RUs, i.e. first and last names, four emphasis contexts are considered for conditioning the VPM: no emphasis (None), pre-emphasis (EMp) on the first of the two RUs, on-emphasis (EM) on the second RU, and post-emphasis continuation (EMc). From the plot we can see that the VPM has successfully captured the increase in amplitude of the pitch movement in EM, as identified in literature~\citep{liu2005parallel}. Moreover, both results we obtained in modelling prominence are in line with our findings with the WSFC~\citep{gerazov2018wsfc}, thus, validating our third hypothesis that the VPM is able to model prominence.

\begin{figure}[t]
  \setlength\belowcaptionskip{-10pt}
  \centering
  \includegraphics[width=.8\columnwidth]{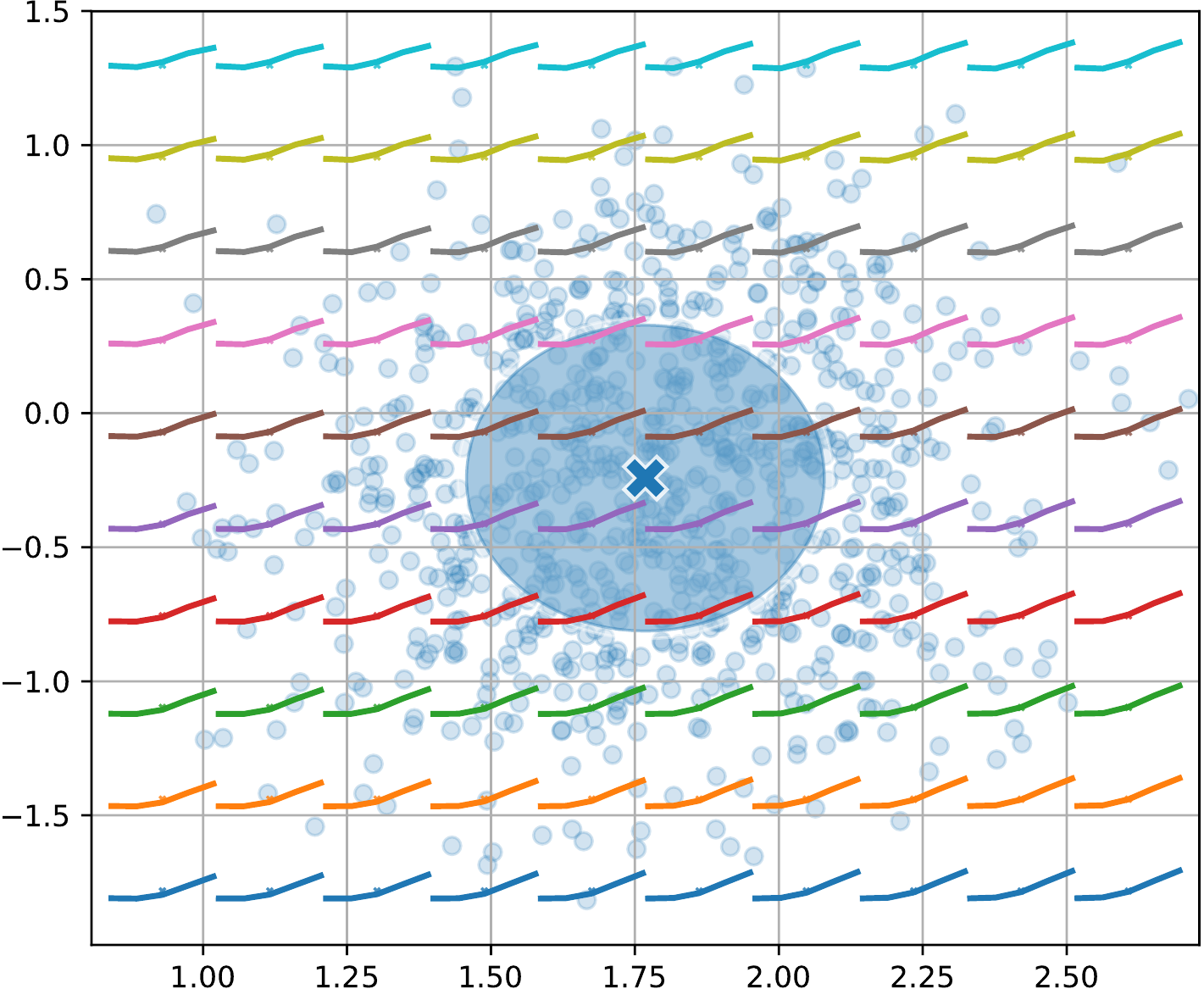}
  \caption{Sampling (blue circles) around the input mapping (X) in the prosodic latent space of Tone 2 obtained with the VPM-VCG for \small{\tx{Chen}}.
The latent space does capture variation in the contour's shape, albeit a small one.
Note that the mapping is not centred in the latent space due to the use of a small regularisation coefficient $\lambda$, see Fig.~\ref{fig:reg} for comparison.}
  \label{fig:sampling}
\end{figure}

\begin{figure}[t]
  \setlength\belowcaptionskip{0pt}
    \centering
  \includegraphics[width=.95\columnwidth]{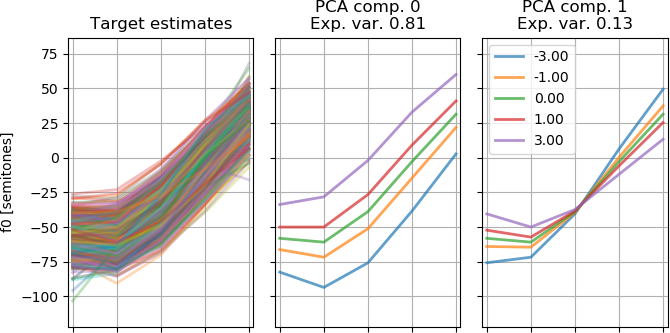}
 \vspace{5pt}

  \includegraphics[width=.95\columnwidth]{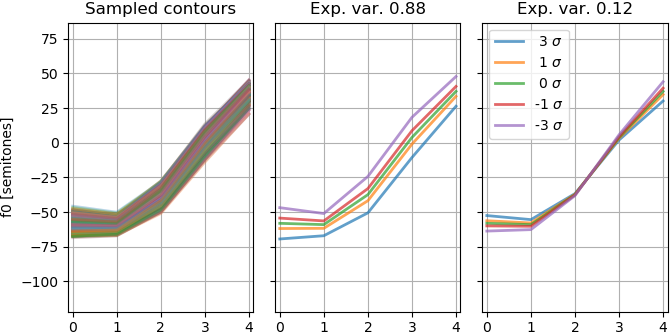}
  \caption{Final analysis-by-synthesis targets for Tone 2 in \small{\tx{Chen}} from the SFC (top row), and contours generated by sampling the latent space with the VPM\nobreakdash-VCG (bottom row), with the variation explained by the first (centre column) and second (right column) PCA components for both sets of contours.
The VPM successfully captures the different components of variation in the contour's shape.
The amplitude difference is due to the approximative nature of the SFC targets, as well as the local averaging effect dependent on random sampling of the latent space during VPM training.}
  \label{fig:var}
\end{figure}

\subsection{Exploring variation}
\label{sec:var}

What is more interesting in Fig.~\ref{fig:emph}, as well as in the decomposition shown in the bottom left plot of Fig.~\ref{fig:example}, is that the VPM has managed to capture a variation of the prototype shape of Tone 3 conditioned on emphasis. Specifically, we can see that post-emphasis the contour's slope flattens out, reflecting reduced pitch dynamics in line with post-focus compression~\citep{xu1999effects}.
Moreover, Tone 3 can be seen to transform from its usual low tone into a rising tone similar to Tone 2 in the first focused RU preceding the second focused RU (pre-emphasis). Since this data is made up of single tone utterances, the captured change in shape is in fact the Chinese tonal sandhi, i.e. a phenomenon in which a Tone 3 preceding another Tone 3 changes into a Tone 2 \citep{xu1997contextual}. These observations confirm the added value brought by the variational encoding scheme and our fourth hypothesis that the VPM can model context-specific variation in the prosodic prototypes.
% We suspect that this anticipatory effect is due to tone sandhi within an emphasis context, however its detailed analysis is outside of the scope of this paper. We believe that it is the ability of the VPM to capture these large contextual changes in the tone shapes that leads to the observed superior performance.

% \GB{We need to rework this section!!!}

%\GBR{The fact that we rely on the function's context to structure the latent space and model the function contour's variation, limits the ability of the VPM to capture shape variations within a specific context. Namely, during training, this process relies on the sampling of the latent space around a context (and RU position) mapping, allowing for succeeding samples within the same vicinity to produce a localised averaging effect.}{
Since the variational encoder of the VPM is fed by the function's context, its latent space is made to capture variations that can be explained by our contextual features.
These features do not entirely reflect the structure of the manifold encoded in the prosodic shapes. %, e.g. see Fig.~\ref{fig:sampling}.
% Even with this limitation, the VPM is still capable of capturing context-specific, as well as within context variation, as we show in Sec.~\ref{sec:var}.
%\GBR{
However, due to the random sampling around the variational encoder mappings in the latent space during training, the VPM has the potential to capture a part of the within context variation.

To evaluate this, we trained single scope, i.e. one RU long, tone prototype contours with no context.
These contour generators effectively see a single input feature combination, which their variational encoders map to a single point in the prosodic latent space.
We now sample around this point with a normal distribution using the output standard deviation by the encoder, as shown for Tone 2 in Fig.~\ref{fig:sampling}, and generate prosodic contours, shown in the bottom left corner plot in Fig.~\ref{fig:var}.
%} {Too much complicated. If you suppress context, why not using a variational AUTOencoder? that should capture not only what is correlated with context but with style components of the raw prosodic signals...}

For comparison, we use the final targets obtained in the last step of the analysis-by-synthesis loop for the SFC, shown for Tone 2 from \tx{Chen} in the top left plot in Fig.~\ref{fig:var}.
Note here that these final targets are not perfect representations of the prototypical contours themselves, as coinciding contours could jeopardise the error distribution in the SFC.

Fig.~\ref{fig:var} also shows the variation captured with the first two components of a Principal Component Analysis (PCA) based decomposition for multiples of the standard deviation in the transform domain.
We can see that the target contours exhibit a variation of 81\% that can be explained by the first PCA component, 13\% by the second, and 6\% by the rest of the PCA components.
%i.e. by simple amplitude scaling. In addition, there is a variation in the target contours of 6\% that cannot be explained by the first two PCA components.
On the other hand, 88\% of the variation captured with the VPM can be explained by the first PCA component, and 12\% by the second.
Since we have limited the prosodic latent space to two dimensions, the VPM cannot capture the variation in the data explained by the higher PCA components.

These results reaffirm the VPM's ability to capture prototype shape variations beyond simple amplitude scaling, compared to the WSFC model.
Moreover, we can see that even in the absence of context based conditioning of the latent space, the VPM still manages to capture a part of the variation in the data through the random sampling process.
However, this process has its limitations that take the form of a localised averaging effect.
Namely, during training, succeeding samples are mapped randomly in the latent space, thus different contour shapes may be mapped within each other's vicinity that will result with the model learning an average contour between them.
The effects of this averaging can be seen in the reduced amplitude of the VPM generated contours in Fig.~\ref{fig:var}.
Nevertheless, these results confirm our final hypothesis.
% }{Not convincing: the SFC gathers more variance with just a linear decomposition!}

% In fact, the amount of variation captured by the VPM compared to the target contours is due to the suboptimal variation capturing mechanism within a specific context based on random sampling.

\section{A unified prosodic latent space}
% \GB{Too sketchy: describe how you group functions. What the latent space is supposed to encode now? What is the benefit of this grouping? Why the grouped latent space is still with 2 dimensions?}
% The mapped prosodic latent space of context-specific variation of the prosodic prototypes can be extended into a ``unified'' prosodic latent space across the different functions.
% This might lead to insights of prototype shape distribution within one level of the prosodic hierarchy.
% For instance, one can think of an attitude prosodic space, or a syntactic prosodic space, or a morphological prosodic space.
% The VPM is amenable to such an exploration as we demonstrate with the attitude prosodic latent space shown in Fig.~\ref{fig:unified} obtained by representing all attitude prototypes with a single contour generator that is conditioned on the attitude type.

\begin{figure}[t]
  \setlength\belowcaptionskip{-10pt}
    \centering
  \includegraphics[width=.98\linewidth]{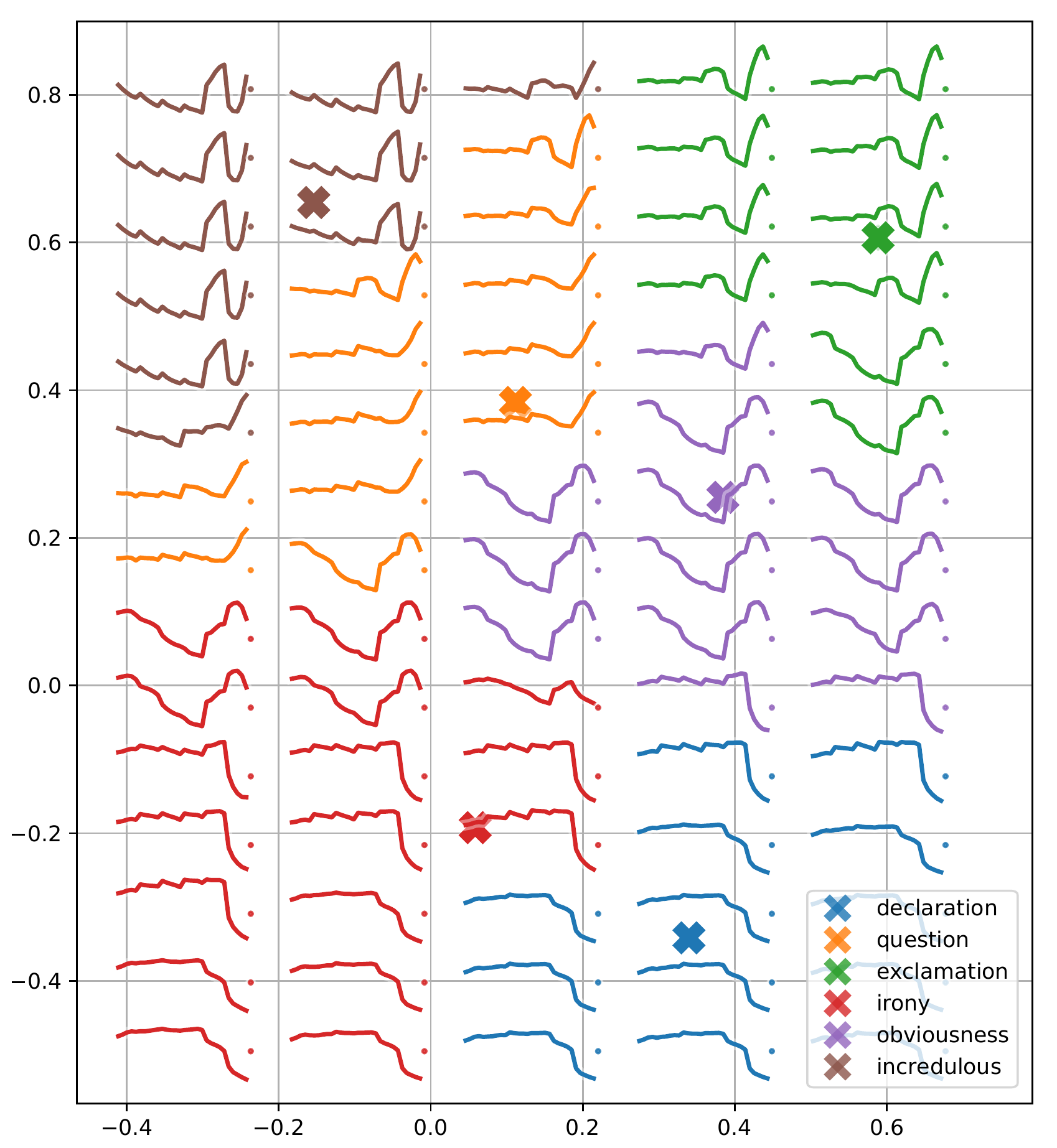}
  \caption{Attitude prosodic latent space obtained with the VPM-VRCG for the \small{\tx{Morlec}} database showing the 6 attitudes present in the data: declaration (DC), question (QS), exclamation (EX), suspicious irony (SC), obviousness (EV), and incredulous question (DI).}
  \label{fig:unified}
\end{figure}

The contour generators in the VPM can also be extended to encode different function prototypes within a single ``unified'' prosodic latent space.
For instance, one can think of mapping an attitude prosodic space that will encompass all of the different attitude functions in the data, or a syntactic prosodic space with the different dependency functions, or a morphological prosodic space that would encode tone and focus for example.
This might lead to insights of prototype shape distribution within one level of the prosodic hierarchy, and might be closer to how humans structure prosody.
We give here only a demonstration of this concept in the form of an attitude prosodic latent space shown in Fig.~\ref{fig:unified}.
The latent space is obtained by representing all attitude prototypes with a single contour generator that is conditioned on the attitude type.

\section{Conclusions}
The proposed Variational Prosody Model uses a deep architecture of variational (recurrent) contour generators to decompose the prosody into its underlying \emph{clich{\'e}} prototype contours and captures a well structured prosodic latent space of their context-specific variation. The VPM plausibility has been demonstrated across two very different languages, and it has been shown to outperform previous state-of-the-art decompositional prosody models, and perform on a par to standard deep models. The prime contribution of the VPM is its incorporation of variational encoding that has been shown as beneficial towards the exploration of the underlying context-specific spatiotemporal variation of the constituent prototype contours in prosody.
Another prospective use of the VPM, that has not been explored in this paper, is its application in speech synthesis, where we believe that the modelled variability can be used to generate a more dynamic and natural prosody that is not as affected by averaging effects.

\section{Acknowledgements}
This project has received funding from the European Union's Horizon 2020 research and innovation programme under the Marie Sk\l{}odowska-Curie grant agreement No 745802 ``ProsoDeep: Deep understanding and modelling of the hierarchical structure of Prosody''.\footnote{\url{https://gerazov.github.io/prosodeep}}

\bibliography{refs}
\end{document}